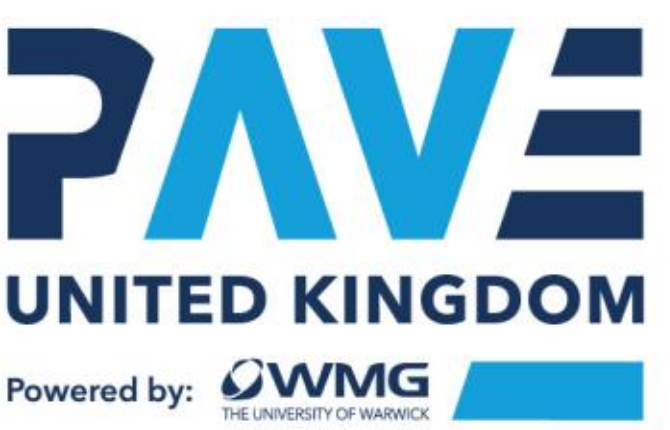

# SELF-DRIVING TECHNOLOGIES NEED THE HELP OF THE PUBLIC

## A narrative review of the evidence

**Jonathan Smith**
**Prof Siddartha Khastgir**

## About PAVE United Kingdom

Partners for Automated Vehicle Education United Kingdom (PAVE UK) is a national initiative that aims to enable public trust and acceptance of self-driving technology through inclusive and accessible public awareness and education programmes.

Founded by the Department of Business and Trade, the Department for Transport, the Centre for Connected and Autonomous Vehicles, Transport for West Midlands and WMG, the University of Warwick, PAVE UK believes future users are the core of this new technology and help shape a safe self-driving future in the UK.

## PAVE United Kingdom founding partners

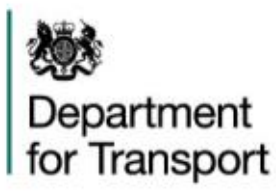

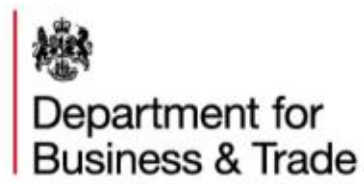

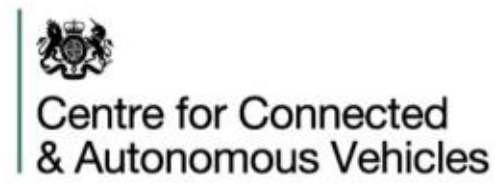

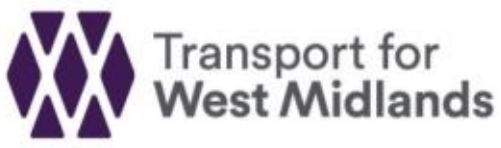

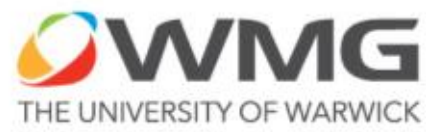

For more information about this report or PAVE UK please contact paveuk@warwick.ac.uk or visit www.paveuk.org

# Executive Summary

History has a long tail of redundant technologies that were technically very impressive but designed without adequate consideration and involvement of the public and stakeholders. If public trust is lost in a new technology early in its life cycle, it can take much more time for the benefits of that technology to be realised.

Eventually tens-of-millions of people will collectively have the power to **determine self-driving technology success or failure driven by their perception of risk, data handling, safety, governance, accountability, benefits to their life and more.** The public will be the ones that normalise self-driving technology and **justify its continued investment** from both private and public sources. As the UK Governments 'Connected & Automated Mobility (CAM) 2025' strategy states 'Working with the public is essential to ensure that the technologies developed meet people's needs: removing barriers to accessing transport (e.g. availability and accessibility) and ensuring that the systems deployed take people and goods where they need to go' [1].

**Self-driving technology needs the help of the public to realise its potential.**

Bringing the public on the journey every step of the way will be critical. Reaching such scale requires that we build upon state-of-the art methodologies, historic case studies, and research evidence. This paper will take a narrative approach to exploring significant attributes of trust and engagement with the intention of **developing a model of self-driving technology engagement which will be implemented through the PAVE UK programme.**

We show it is 'not in the good will of humans or our own feelings that will give rise to trust in self-driving technology, but rather in **the information and reasons the public and stakeholders have access to concerning the operation of these technologies'** [2].

**If a mismatch between the public's perception and expectations about the capability of the automated system emerge** and the designers make assumptions about what knowledge the public have, it can **lead to misuse (due to mistrust), disuse (due to distrust) or abuse** of the automated system [3] [4]. This applies to both the basic physical safety and security of the public, [5] and the ability of the vehicle to achieve the goals it sets out to benefit them.

It goes without saying **not all members of the public and all stakeholders will trust or accept self-driving technology**. As we find through exploring other safety-critical technology case studies these differing views can often playout in very public ways such as in the media or through the courts. **These exchanges are engagement itself. Of critical importance is how these exchanges unfold over time**, having substantial potential to advance the industry rapidly if undertaken in good faith and driven by data-driven evidence. The ability of such a direction of travel to be created over time is dependent upon establishing and maintain a trustworthy system: constituting formal and informal Governance mechanisms that adequately include the public and its stakeholders. Importantly we find that it is not necessary that all members of the public and stakeholders agree, but the characteristics of a 'trustworthy' system that lays the foundation for sustainable success.

Orientating engagement around **what matters to the public** creates the potential for ever more sophisticated conversations, greater trust, and moving the public into a progressively more active role of critique and advocacy. Making self-driving technology 'relatable' to an individual's circumstance will enhance engagement. We find from the evidence that industrial **experts** **often misunderstand what matters to the public, users, and stakeholders**. The public have a deficit of knowledge, and that **lack of knowledge can drive rejection** [6] and efforts should be made to 'level the social and cognitive authority' of experts [7]. Conversely a key characteristic highlighted within the paper is that the public have a surprising capacity to digest complex information leading to acting as informed critiques and advocates when highly engaged [8].

Consequently, there is a 'necessity' of regularly calibrating the public's knowledge and expectation of autonomous vehicles through educational campaigns and legislative measures mandating user training and timely disclosure from car manufacturers/developers regarding their product capabilities' [9].

The implementation of such characteristics can slow down the processes of innovation in the short term and even feel quite uncomfortable, but they are nevertheless critical steps in a systems-of-systems approach to successfully implementing technology. Thus creating a 'trustworthy' basis on which to demonstrate technology has been developed for the right reasons and in the right ways.

As increasingly large volumes of information become available regarding self-driving, we have concluded that **engagement programmes must develop approaches to defining the right information at the right time (in the right format).**

Self-driving technologies are an emerging technology and we, are on a journey towards them. At the level of each individual they are on their own journey towards first contact with the concept of self-driving technology, through knowledge acquisition, maturing to actual interaction or use over time. Reflected in the fact that trust is dynamic, as a result of continued knowledge acquisition (and often old knowledge forgotten [11]) **a model of self-driving technology knowledge acquisition would be useful.** We conclude such a model would provide guidance on knowledge attributes through which engagement programmes can effectively hook into mental models as they form for those who want, need, or care about the information [12]. The paper concludes by proposing such a model, building on the evidence found.

Whilst lived experience of self-driving technology will be an important component of engagement in the future **our model understands that trust creation begins much before the first interactions between a user and a self-driving technology** [13] **and refines an engagement approach that identifies those that want, need, or would benefit from engagement at this time** [12]. Through systematic analysis of **over 100 studies on self-driving** trust, acceptance and adoption a mode of engagement and education, centred on what matter to the public, is proposed.

This model is coupled with strategies, tactics and tools for the implementation of that model. Finding that engagement with the public and stakeholders in the 'creation' stage of educational material creates a foundation for scalability. **It's important to emphasise that this is not a case of the public defining independently what matters but structuring the start of engagement around the public's perspective** [14] **providing a platform for deepening the discussion between the public and domain experts and building a framework of scalability that works on both sides.** Stakeholder engagement is a critical aspect of effective engagement so continued and updated understanding and involvement of stakeholders is essential [15]. Extending this to a complex safety critical eco-system the role of stakeholder engagement includes addressing dynamic relations and complex environments that utilises different frames of reference, networks, specialist insights, and ongoing changes [15].

Through a narrative review of the evidence this paper makes the case for a UK national programme of **scalable inclusive and accurate education and engagement**, that will enable trust through learning, communicating and engaging. The paper draws actionable lessons from the evidence that would affect such a programme.

# 1. Introduction: self-driving technologies need the help of the public

Self-driving technologies are approaching fast, and they need the help of the public. Since 2015 the UK Government has invested £300m in the development of self-driving technology and has committed a further £150m [1]. UK industry has responded to the call by far exceeding this Government investment, with an estimated £480m invested between 2018-22 creating 1,500 new jobs [16]. As a result, a flourishing eco system of small and large organisations across a range of technology and capability has evolved. This momentum has been confirmed by the UK having attracted the largest ever AI-start up investment in Europe through self-driving technology, with Wayve technologies securing £1.1 billion [17]. At the same time the Automated Vehicles Act 2024 (AV Act) has passed through parliament paving the way for the safe scaling of self-driving technology on UK roads [18]. An end-to-end capability of technology testing has been established from virtual testing, to track testing and real-world testing in the creation of 'CAM-Testbed-UK'. On-road trials are accelerating learning from Edinburgh to Birmingham to London [19], [20]. Looking forward the introduction of self-driving technology is forecast to create over 38,000 new jobs and add £42 billion to the UK economy [1]. Road safety is a significant concern nationally (1,766 fatalities and 28,941 serious injuries in 2022) and international with the United Nations and the World Health Organisations having launched dedicated road safety programmes in 2020 and 2021 [21], [22]. Self-driving technology offers the potential to significantly improve road safety. Finally possible societal benefits are considerable: from better integrating rural communities, reducing isolation for older people and those with disabilities, to helping deliver essential goods and improving education, work and leisure [1].

The technical challenge of developing reliable new technology is significant, however there is more to the successful introduction of technology. History has a long tail of redundant technologies that were technically very impressive but designed without adequate consideration and involvement of the public and users. If public trust is lost in a new technology early in its life cycle, it can take much more time for the benefits of that technology to be realised. This has been evident from the recent experience of Smart Motorways in the UK, which following a series of incidents lost the support of public sentiment and were eventually halted in UK policy, despite encouraging technical data on their effectiveness [23].

Technology success doesn't exist in a vacuum but has a complex relationship with people and policy. In line with this the importance of these challenges are clearly recognised in UK Policy. The UK Governments 'Connected & Automated Mobility (CAM) 2025' strategy states that 'Working with the public is essential to ensure that the technologies developed meet people's needs: removing barriers to accessing transport (e.g. availability and accessibility) and ensuring that the systems deployed take people and goods where they need to go. Otherwise, trust in the technologies could be impacted, damaging the potential for CAM more broadly' [1]. The Government has indicated it will take this commitment seriously illustrated by the Automated Vehicle Act which has incorporated recommendations by the Law commission to make misrepresentation of automated vehicle capabilities a 'criminal offence', the requirement for the Safety Principals to include consultation with road users, and making the role of the 'user-in-charge' clear, among others [24][25].

Early policy and research work has created a baseline understanding including the 'Driverless Futures' project [26] 'Future of Transport Deliberative Research'[27] and the 'Great Self-driving Exploration' [28]. These projects have developed valuable UK insight such as the need to address safety, accessibility, and providing the public with a basic understanding of what self-driving technology is. However, these projects reached a very small proportion of the public and the rate of change the technology is experiencing requires us to both widen the net and match the current pace of technology change. The public should not be sampled, but inclusively and continuously engaged in the processes of technology development. The public will be the ones that normalise self-driving technology and in many cases be the users that justify its continued investment from both private and public sources. Self-driving technology needs the help of the public to realise its potential.

It is necessary for Government and industry to prepare for this wide scale awareness of self-driving technology which will lead to the forming of conceptions, opinions, acceptance (or rejection), and adoption. This will eventually be tens-of-millions of people who will collectively have the power to determine self-driving technology or failure. Driven by their perception of risk, data handling, safety, governance, accountability and more.

Bringing the public on the journey every step of the way will be critical. Reaching such scale requires that we build upon state-of-the art methodologies, historic case studies, and research evidence. This paper will take a narrative approach to exploring significant attributes of trust and engagement with the intention of developing a model [6] of inclusive and accessible self-driving technology engagement. First, we will set a theoretical grounding in trust which we will explore through the lens of two cases studies (big data in healthcare and nuclear power policy). Having illustrated the role of 'engagement' in the building of trust we will explore the key pillars of building successful 'engagement' campaigns. Moving onto grounding both of these discussions (trust and engagement) in self-driving technology itself by analysing insights to what aspects of self-driving technology users consider important and engaging. The paper will conclude by building on the learning developed throughout to propose a model of implementation for inclusive and accessible self-driving technology engagement which will be applied to the activities and strategies of PAVE United Kingdoms activities.

# 2. Trust to trustworthiness: a systems of systems approach

## Why trust?

This paper aims to develop a model for the accurate and inclusive engagement and education of the public in self-driving technology. To productively arrive at this outcome we must first ground ourselves in the correct theoretical concept(s). Trust has a long history of study going back to Greek philosophy and has proven to be foundational in understanding politics, organisations, brand building, marketing, personal relationships, product design and technology adoption among others. In the broad context of technology and the specific context of self-driving technology a wealth of literature finds that trust is both important and complex, going well beyond the tangibility of the technology itself. Trust is a system-of-systems linking technology, people and policy. This phase of the paper illustrates that time and again accurate and inclusive engagement is directly linked to trust. Enabling engagement in technology requires an understanding of how this trust building system forms.

To illustrate this capacity for trust to influence the trajectory of technology lets first explore trust through two recent cases studies, the first, big data in healthcare and second, nuclear power. Both case studies reflect important characteristics of self-driving technology in that they involve complex emerging technologies in which industry and society are learning in parallel. They are safety critical where mistakes will result in significant adverse personal, public, and environmental impacts, and they are industries that require the tri-pillars of private investment, Government policy (and investment), and public acceptance to realise their potential. Learning from these examples will emphasise why trust is important, and begin to provide insight to how on the journey to realising the potential benefits of technology a system of trust forms (or fails to form).

## Exploring trust through big-data and healthcare: case study one

Data will bring benefits to all parts of health and social care [29]. Two prestigious organisations, the NHS and Google DeepMind began work on this challenge in 2015, shortly after Google had acquired DeepMind and before GDPR was introduced. A public announcement outlined the scope of the relationship with the London Royal Free NHS trust to help hospital staff detect and diagnose when patients were at risk of developing acute kidney injury [30]. However, in 2017 The New Scientist obtained the data-sharing agreement in place for the project revealing that all data (including 5 years of medical history) for the 1.6 million patients that passed through the trust each year had been shared [31]. The data shared went beyond the scope of the project and the patients only recourse was via a complex opt-out mechanism. A subsequent legal case was brought representing the patients. The case was reviewed by the Information Commissioners Office (ICO) finding that the UK Data Protection Act was not complied with and that the patients would not have reasonably expected their information to be used in this way [32].

We might find it hard to make a case *against* improving detection and prevention of kidney disease, yet this case attracted criticism in the media (including the BBC [30], Financial Times [32], New Scientist [31], and CNBC in the US [33]) interrogation from academia [34], and a warning of how not to handle data from the National Data Guardian, Dame Caldicott [35]. How did a negative narrative emerge around a project with good intentions? Its notable that, in The New Scientist article that originally broke the story immediate responses from the NHS trust and DeepMind are vague including '[The NHS] did not respond to questions about what opt-out mechanisms are available to its patients'. This suggests the potential for concern over the use of patient's data had not been considered or were of a low priority, possiblly reflected in the relative perspectives, power and knowledge advantages of the decision makers [36]. The NHS could and should have been far more transparent with patients as to what was happening [32] despite the well-intentioned ends.

Whilst it is unrealistic to expect all stakeholders to agree, the lack of effort in the processes itself to give all stakeholders a voice to influence decision making and the lack of transparency in making critical decision with patients' data made it difficult for the project to justify itself as 'trustworthy' once the debate moved into the public sphere.

As details of the case became public a void emerged between how the project had been described, the specific aim of improving detection and monitoring of kidney disease, to what was happening in practice, the sharing of all data from all patients. Some perfectly justifiable reasons for this emerged later, including that patient data is not siloed according to treatment meaning part of the research project was focused on how to segment relevant data [31]. However, both making these claims after a breach of trust and been unable to show how the mechanism of the data-sharing agreement accounted for them was all too little too late.

The media attention and public scrutiny in this case-study undoubtedly fuelled a sense of public mistrust. This mis trust was reflected in the project, the organisations involved and more widely in the application of data to advance our healthcare system. However, the building blocks of mistrust were in place well before the case became public. Technology 'must advance in a way that meets and exceeds existing regulatory frameworks and societal expectations' [34] by adopting fundamental characteristics of transparency, openness and accountability [6]. The implementation of such characteristics can slow down the processes of innovation in the short term and even feel quite uncomfortable, but they are nevertheless critical steps in a systems-of-systems approach to successfully implementing technology – both leading to added value in the development of the project and a 'trustworthy' basis on which to demonstrate technology has been developed for the right reasons and in the right ways.

We can well imagine an alternative scenario where a sub-set of patients were engaged and consulted in the research project leading to the verification (and/ or adjustment) of difficult data-management decisions by the stakeholders that have an invested interest. In such a scenario unexpected value can be derived as members of the public have a surprising capacity to digest complex information and make nuanced judgements when engaged in the right ways [2]. Trust is not absolute but a critical mass of negative and positive consideration [13]. Stakeholders must adhere to explicit and transparent principles of good practice [6] so individuals can arrive at their own level of trust judgement, what Rempel et al call calibrated trust [6].

Powles and Hodson (2017) conclude that 'the failure on both sides [DeepMind and the NHS] to engage in any conversation with patients and citizens is inexcusable' especially considering recommendations made in two prior reports, the Caldicott review and the Nuffield Council report on bioethics, triggered by the failure of an earlier 'care.data' scheme [34]. The power of decision making was entirely with the organisations involved. The relative weak position of patients legally and socially was not accounted for until firstly investigative journalism and secondly the concluding of a lengthy and costly legal battle. Whilst the system of 'legal' Governance eventually prevailed, this was not the optimal route to the building of trust and realising the potential economic and societal benefits from the technology. This raises the question of how to inspire a system of self-governed trust that reaches beyond the minimum legal requirements, which if done well unlocks hidden value?

## Exploring trust through nuclear power policy: case study two

The second case that will enhance our understanding of trust is from the nuclear energy industry. Nuclear energy has been a contentious point of debate for decades with negative and positive links to environmental goals, links to defence, and decades long life cycles that have an impact across multiple generations. After a 2003 policy commitment to ensure the 'fullest public consultation' before proceeding with the building of new nuclear power stations [37] it was found in a 2006 Judicial Review that the '[public] consultation was flawed' [37] in establishing new nuclear policy. Leading to a second extensive public consultation.

As the prior case study illustrated the risk of not engaging, we will benefit from understanding the practicalities, nuances, difficulties and outcomes of this second comprehensive consultation on how to engage, especially in this case which dealt both with a challenging technology and was conducted in a challenging environment following the judicial review. Due to the intense scrutiny by multiple parties we can draw upon significant documentation including the independent evaluation report commissioned at the time, which found overwhelming support for adequate or better levels of good practice in the second consultation [37]. This included written and online consultation, stakeholder events, and public engagement – all prepared with a Citizen Advisory Board.

Following a 2006 update to UK Government policy that concluded 'new nuclear power stations would make a significant contribution to meeting our energy policy goals' Greenpeace applied for a judicial review of the decision on the grounds the [public] consultation was flawed [24]. The challenge was upheld on the basis of an 'overriding need for fairness in any consultation process' [37] leading to a second consultation.

The Judicial review was brought by an influential stakeholder, Greenpeace. Their work at this stage of the processes triggered a second consultation that enabled all stakeholders to have a greater contribution to the forming of Government policy.

However, after their successful challenge of the first consultation and subsequent involvement in the design of the second, on the morning of stakeholder deliberations held across the UK Greenpeace (and Friends of the Earth) pulled out of the process, making front page news. This continued with some serious exchanges undertaken through the course of the consultation. Along with national media attention, the Nuclear Consultation Working Group published an extensive paper criticising the process and Greenpeace brought a further legal case against one of the organisations delivering the deliberative workshops.

Differences between the first legal case by Greenpeace and the second can be seen. Where the first was as an outsider (and could be seen as representing all stakeholder groups) challenging an unfair process, the second relates to the challenge of specific detail in the processes (and a *choice* to pull out). The power and influence of Greenpeace as one stakeholder should be noted here – used to effect in the first instance and potentially as a force of disruption in the second.

Whilst serious in nature and doubtless difficult for the individuals involved on all sides these exchanges are '*engagement'* itself. Power and influence are unavoidable between stakeholders and attempts to level the playing field support the building of trust [7].

The deliberations went ahead with a broad range of other stakeholder groups, the report noting 'Greenpeace were far from the only group opposing nuclear energy' and 'green NGO's were one of many voices .. [that had] considerable knowledge of the issues' [37].

One of the most powerful forms of engagement is embracing sites of contention as sites of engagement [6] which is most effective if embedded in the processes. The effective handling of shifting power dynamics in the second consultation appears to have led to outcomes of a trustworthy nature, such as new policy proposals on waste management and re-processing which policy makers had not expected to be part of the policy framework until they heard first hand strong feelings expressed in the consultation [37]. No doubt all stakeholders will not agree with the outcome, indeed the independent evaluation notes that of the five issues identified through the consultation, four were explicitly tackled, but one (the role of private sector ownership and management) was not. The case illuminates the fact that strongly opposing opinions can exists alongside engagement also *happening* even when that engagement is serious and protracted i.e. legal proceedings. Moreover, final decision making that it not aligned with some stakeholders can still be formed in an atmosphere of trust if an authentic and influential form of engagement formed a substantial pillar of that decision making. This places emphasis on the role of processes as opposed to outcome, or trustworthy behaviour.

## Trustworthy behaviour

The system-of-systems in which safety-critical technology establishes is, by definition, multifaceted. Trustworthy behaviour therefore cannot be established by a given individual or organisation, but as a defining characteristic of the system. *Figure 1: Regulation focused model of trust building and public engagement (Winfield et al 2018)* represents a trust building technology Governance system, ethical (purple), technical (blue) and regulation (red) intersect in complex ways where a mix of stakeholders combine to form each block [38]. For example technical methodologies for verification and validation must include developers, insurance, academic research, test houses, road operators, vehicle manufactures, and so on. However, the figure shows that in building trust this technical block (blue) should build upon ethical inputs.

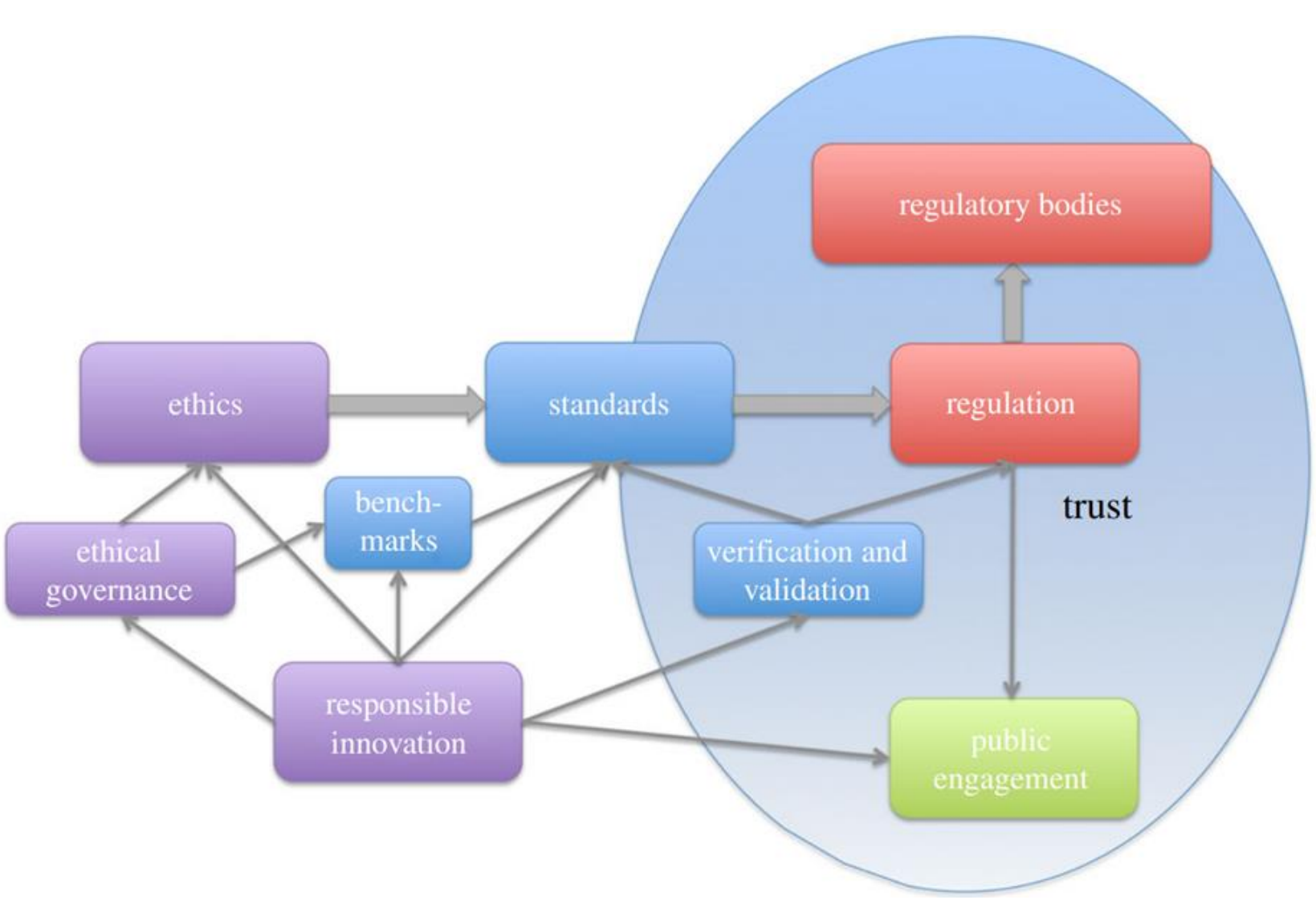

*Figure 1: Regulation focused model of trust building and public engagement (Winfield et al 2018)*

Rempel et al (2018) argue that transparency is the key pillar of trustworthy behaviour. Defined as 'public openness' such as releasing data sets, meeting minutes and engaging with stakeholders [6]. When one stakeholder observes such behaviour among its counterparts, especially with value adding benefits, such behaviours become embedded as the norm re-enforcing one another over time. Scrutiny of information that has become public gives rise to Government and industry becoming *accountable* for good practice in technology and innovation [6]. As was observed in the Nuclear case study such openness can lead to serious and uncomfortable engagements, such as legal action. Parallels are emerging in the self-driving technology, for example the release of data which have stimulated a debate among stakeholders [39]. These exchanges are engagement itself. Of critical importance is how these exchanges unfold over time, having substantial potential to advance the industry rapidly if undertaken in good faith and driven by data-driven evidence. The ability of such a direction of travel to be established over time is dependent upon establishing and maintain a trustworthy system: constituting formal and informal Governance mechanisms. Grounding this discussion in the objectives of this paper, what is the role of the public and engagement.

The final (green) block of Figure 1: Regulation focused model of trust building and public engagement (Winfield et al 2018) shows that 'public engagement' is central to this trust building system. Transparency is not a replacement for public engagement but facilitates successful engagement which acts as a re-enforcing mechanism for the building of a system of trust [6]. Let us now turn our attention to trust in the technology of self-driving technology. It is only by considering both the system of Governance and the Technology System that we will arrive at a complete understanding of trust.

## Empowering user trust in self-driving technology

It is important to define trust in the self-driving technology context. In order to explore trust, we adapt the technology definition of trust from (Lee and See, 2004) as, "a history dependant attitude that an agent will help achieve an individual's goals in a situation characterised by uncertainty and vulnerability" [40]. The addition of the reference to "history dependant" is particularly important for this work because prior knowledge about the system's capabilities and limitations affects an individual's attitude towards a system, thus affecting their trust. Trust is said to be influenced by various factors. Lee and See (2004) suggest specific design, evaluation, and training consideration for designing trustworthy systems which are, showing past performance, comprehensible evidence of system performance, simplification and understanding, purpose, user training on reliability and Governance, and anthropomorphising of the automation [40]. Work conducted by Khastgir et al (2018) suggests trust in the context of increased vehicle automation can include knowledge, certification, situation awareness, workload, self-confidence, experience, consequence and willingness [4]. Similarly, Chilson (2022) in the specific context of self-driving technology considers repeatability, predictability, reliability, transparency, reconstruct ability, and explicability to be key concepts [2].

In contrast to the slightly more abstract 'system of system' perspective explored till now these findings effectively bring the user into focus when understanding trust. Indeed, it's possible we only gain a complete understanding of trust at the level of an individual. This perspective of the individual is an important consideration as it is not in the good will of humans or our own feelings that will give rise to trust in self-driving technology, but rather in the information and reasons we have access to concerning the operation of these technologies' [2]. Recognising the link between engagement and education programmes at the heart of this paper, and this 'information and reasons' it is necessary that we explore further how such information and reason forms at the level of the individual and the technology.

## Informed Safety

An important section of the literature on trust specifically focuses on the trust formation stage whereby the self-driving technology performance, capabilities and interaction quality are the most important trust building paths [41] [42]. These core factors manifest at two levels, the first regarding basic physical safety and security, [5] and the second regarding the ability of the vehicle to achieve the goals it sets out to benefit the user. However, the interaction between these two elements can be nuanced and discreet. In both instances the prior expectation of the user sets a critical benchmark against which trust forms, for example if the user expects an emergency stop when a pedestrian jumps in front of the vehicle, their trust level is less likely to fall once experiencing an emergency stop. This is a critical concept in the trust research referred to as 'Calibrated trust' [13] which has been further refined to the self-driving technology context by Khastgir et al (2018) as 'Informed Safety' [4]. Informed safety means informing the driver (via static and/or dynamic knowledge) about the safety limits of the automated system and its intention, it is not just about providing rules of usage, it includes the background information, understanding and knowledge about how the system operates [4]. A mismatch between drivers' perception and expectations about the capability of the automated system, and the designers' assumptions can lead to misuse (due to mistrust), disuse (due to distrust) or abuse of the automated system [3] [4].

Consequently, there is a 'necessity of regularly calibrating the public's knowledge and expectation of autonomous vehicles through educational campaigns and legislative measures mandating user training and timely disclosure from car manufacturers/developers regarding their product capabilities' [9] (mirroring the characteristics of transparency and openness identified in 'trustworthy Behaviour'). Khastgir et al (2018) found that for low capability automated systems the introduction of knowledge increased the level of trust in the system from 32.4% to 65.4%) [4], Zhang eta al (2024) found that the role of explainability – the self-driving technology ability to describe the rationale behind their outputs in human-understandable terms had the most profound impact on trust [43], and a number of self-driving technology trials have found that the experience of riding in a self-driving technology (real and simulated) increases the likelihood of using a self-driving technology in the future, especially those with higher levels of anxiety towards use intention [44].

However, in some instances too much knowledge is not an appropriate solution and can lead to dis-use or misuse, therefore there is likely an optimum match between a given systems capability, the knowledge provided about that specific system, and the prior knowledge of the specific user to optimise trust [4]. Studies reflect this nuance. In conducting an analytical literature review and in-depth interviews with experts Valentine et al (2021) conclude that trust creation begins much before the first interactions between a user and a self-driving technology [13] and following a self-driving technology simulator study with a 1-week gap in self-driving technology exposure Hunter et al (2022) find that people can forget some of their gained trust (or distrust) in automation over time [11]. Furthermore, Zhang et al (2024) found that after presenting participants with 'six scenarios [they] showed more blame and less trust attributed to autonomous vehicles, despite the scenarios being identical in antecedents and consequences concluding that blame and trust are shaped by 'stereo-typical conceptions'[9].

To bring these findings back to the earlier discussion on 'transparency' we can conclude that all information in all forms is likely not the solution. More mechanisms for accessing information as required so that the right information at the right time (in the right format) can be determined through trustworthy behaviour.

## Trust *with* and trust *in*

As we can see trust remains a diverse concept, even when its specifically applied to and developed within the context of self-driving technology. Khastgir et al (2018) unpack this by differentiating between 'trust *in* the system', and 'trust *with* the system'. Trust *in* the system means the users trust in the capabilities of the system and/or in the system's ability to do what it is supposed to do. Here we are concerned with the formation of knowledge before using or being exposed to the actual technology. Different knowledge types that form this knowledge acquisition journey have been defined as Internal Mental Models (prior beliefs influenced by external sources such as word of mouth, media, marketing), Static (understanding of the functionality of the automated system administered before experiencing), Real-time (automation health, current state of the automation, near-future intentions of the automation which brings the user in-the-loop) [4]. This later knowledge type, 'real-time' signals the shift to trust *with* the system. Trust *with* the system means users awareness or attitude towards the limitations and their subsequent ability to adapt their use of the system to accommodate whilst still deriving the expected benefits [4]. Here we are concerned with the correct and effective use of the system, which can further extend to complex interplay with the user such as during malfunctions and trust recovery.

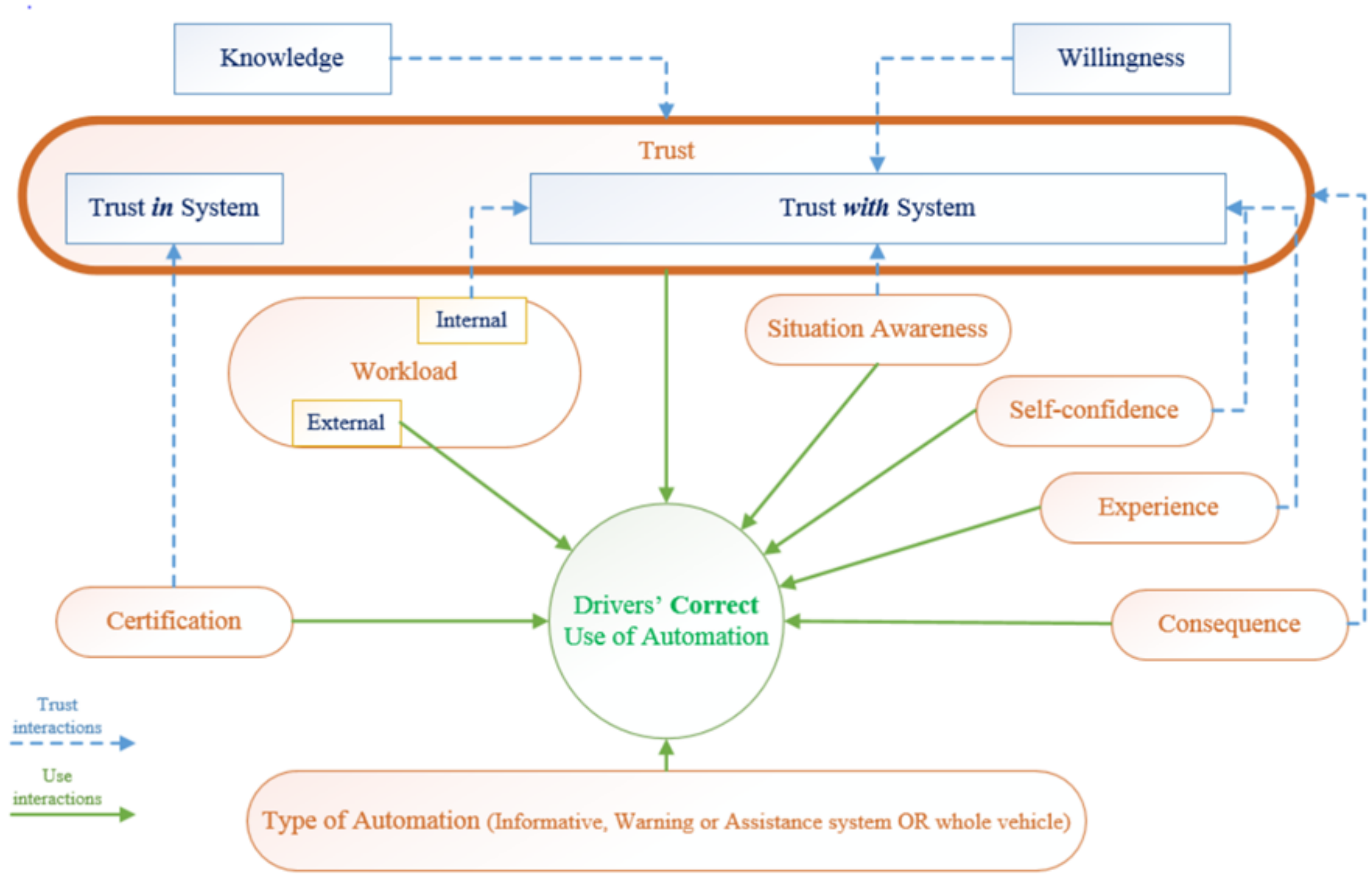

*Figure 2: self-driving technology technology focused model of trust (Khastgir 2019)*

*Figure 2: self-driving technology technology focused model of trust (Khastgir 2019)* captures the building blocks of trust *in* and trust *with*. The role of 'certification' mirrors much of the Governance discussion/ model at the start of this section. The role of 'knowledge' is reflected in the objective of the paper itself, extending knowledge to a wider consideration of engagement and education. Together this model identified these two elements, knowledge and governance, as the central influencers of trust *in*. Both these elements are one step removed from the technology itself suggesting that at least in part their application transcends a specific self-driving technology or brand. The elements of trust *with* have a tighter focus on a specific self-driving technology or brand. Building on the findings of a trustworthy system identified earlier we can see again, this time from the bottom (technology) up approach, that the roles of differing stakeholders across the governance and knowledge domains fit together to collectively build a system of trust.

## Inspiring a system of trust: the role of engagement programmes (section conclusion)

It is necessary that we consolidate the findings so far with a focus on the engagement objectives of this paper, to focus the discussion moving forwards. What is the significance between trust and accurate and inclusive self-driving technology engagement? Beginning with a grounding in trust through the lens of two case studies (big data and healthcare, and nuclear power policy) we have been led to answers.

- **Public engagement is central to a trust building system.**
- All information in all forms is likely not the solution, but mechanisms for accessing information as required so the **right information at the right time (in the right format**) can be determined through trustworthy Behaviour.

- Members of the public have a **surprising capacity to digest complex information** and make nuanced judgments when engaged in the right way.
- One of the most powerful forms of engagement is **embracing sites of contention as sites of engagement**. Final decision making that it not aligned with some stakeholders can still be formed in an atmosphere of trust if an authentic and influential form of engagement constituted a substantial pillar of that decision making. This places emphasis on the role of **processes as opposed to outcome, or trustworthy behaviour**.
- The implementation of trustworthy behaviour **can slow down the processes of research and innovation in the short term** but is critical to long term systems-of-systems success (and has the potential to uncover unexpected value).
- **Power and influence are unavoidable** between stakeholders and attempts to level the playing field support the building of trust and the delivery of successful engagement.
- **Transparency is the key pillar of trustworthy behaviour**. Transparency is not a replacement for public engagement but facilitates successful engagement which acts as a re-enforcing mechanism for the building of a system of trust.
- **Transparency is not all information all the time but must be defined alongside other theories of engagement** such as having the right information, at the right time, in the right format (for the right purpose).
- Transparency leads to scrutiny of information that has become public giving rise to Government and industry **becoming *accountable* for good practice in technology and innovation**.
- The application of a specific self-driving technology lens to the concept of trust has **effectively brought the user into focus**. It is not in the good will of humans or our own feelings that will give rise to trust in self-driving technology, but rather in the information and reasons (including engagement and education material) we have access to concerning the operation of these technologies.
- A mismatch between drivers' perception and expectations about the capability of the automated system, and the designers' assumptions can lead to misuse (due to mistrust), disuse (due to distrust) or abuse of the automated system. Consequently, there is a '**necessity of regularly calibrating the public's knowledge and expectation of autonomous vehicles** through educational campaigns and legislative measures.

We have developed a theoretical grounding in the concept of trust. This in turn has illuminated the value of engagement created in a spirit of trustworthy Behaviour. Building on this orientation we will now undertake a deeper analysis of what forms 'trustworthy' engagement.

# 3. Engagement

## Understanding the nuts and bolts of engagement

Its insufficient to simply justify engagement as necessary (which we established in the previous phase of the paper), our learning will be advanced through an investigation into the methodologies and characteristics of engagement. In this phase of the paper, we will build on the theoretical and move into the applied. An easy statement to make, difficult to implement because engagement is a multi-faced concept that encapsulates a wide range of different societal groups with differing degrees of knowledge background and perceptions. It covers a wide range of different engagement scales from one-to-ones, right through to items of material that could potentially reach thousands of people. We will dig into differing concepts of engagement, how good engagement is grounded in (continuous) evidence, before understanding effective tools in engagement (storytelling, stakeholder engagement, co-creation). To conclude the section, we'll summarise how the public can be enlisted to help. Building on the recognition that an understanding of the user perspective is pivotal will lead us to the final phase of this paper: engaging on what matters to the public.

## Bad vs good engagement

### From communicating to engaging

As we have established earlier in this paper knowledge is a critical element of societal engagement. It therefore follows that such knowledge must be communicated. However, knowledge overload is also in-effective and the methods of how to communicate, what to communicate and when are critical questions.

Over the past decade there has been growing consensus that communications (one-way information flow) is shifting to engagement (two-way information flow) [8], [36], [45]. Such a framing understands Communications as stakeholders' cognitive involvement and community capacity building through relationship building [36] wherein the public and organizations who share mutual interests in salient topics … are aimed at goal attainment, adjustment, and adaptation for both the public and organizations benefit [8]. This approach to public engagement in the context of new technological and social developments is reshaping and reconceptualising communications [45].

Even so whatever labels they go by, communication strategies fall along the continuum between notions of engagement as control and engagement as collaboration [8]. Moreover, our second case study on trust, Nuclear Power Policy, illustrated how a weak engagement strategy (in their case a failed consultation) can in fact lead to worst outcomes. Evidence that in-effective engagement is damaging is further found in the UK's nano technology industry with Groves (2011) finding that ineffective stakeholder engagement led to a fall in trust.

*'policy and business actors.... deeply embedded assumptions about how technological innovation creates the future, tended to place obstacles in the way of turning [engagement] aspiration into reality.* [46].

This supports other findings including the desire of stakeholders to be involved in more dialogic forms of engagement (UK biotechnology) [47] and the UK's responsible innovation framework encouraging public and private research activities to enhance stakeholder engagement [48], [49].

Echoing this our discussion on trust illustrated that engagement can be an uncomfortable process (and is in fact enhanced if sites of contention are embraced). Rasing the question: what factors move us beyond traditional communications into trustworthy engagement?

In the first instance a comprehensive research strategy should be formed. Research makes public relations activities strategic by ensuring that communication is specifically targeted to societal groups who want, need, or care about the information [12]. Without conducting research, public relations are based on experience or instinct, neither of which play large roles in strategic management. An asymmetrical approach seeks to determine through research what the public knows and understands about an issue of importance and these learnings are incorporated to the strategy with the aim of persuading the public. Extending further the symmetrical model undertakes research to increase its understanding, but it doesn't seek to use this information to persuade but as the base for the building of mutual understanding between the public and organisations [50].

This second approach, symmetrical, reaches further into the characteristics of trust we have previously explored, empowering the public in a trustworthy process. Characterised in the opportunity for stakeholders to actively contribute and affect outcomes (not just listened to).

To have our research lead us to successful engagement outcomes we require focus. Research underpins a situational analysis which is a two-paragraph statement that summarises the situation. The first defines the situation using the data collected form your research, and the second should recommend solutions with included reference to the problems, barriers, and difficulties that will be overcome. In a symmetrical model this situational statement acts as a guide for initial engagement and will be continually reviewed and updated based on learning and engagement [50].

## Understanding the stage of the journey

As the journey evolves the characteristics of engagement changes. This is a result of the public having greater knowledge leading to shifts in their mental models. Dhanesh (2017) argues that the degree of engagement of a target public (from passive to active) determines how, at a given point, engagement should be undertaken. If well-constructed the public will progressively become more active (engaged), acting as informed critiques and advocates when highly engaged [8]. Similar findings are found in earlier work by Lee and Kwak (2012) [51]. The model is summarised below (adapted from Dhanesh 2017) [8]:

1. Passive: Informed measured by clicks, views and reads
2. Passive: Disseminated too measured by likes
3. Passive/ active: Broadcasts measured by comments and calls
4. Passive/ active: Converses measured by shares or protests
5. Active: Collaboration measured by critique

6. Active: Participate measured by advocacy

The inclusion of measurable outcomes here is key to a full research cycle. Evaluation is a critical aspect of an engagement campaign such that intended outcomes can be benchmarked and further research data can be collected to underpin a cycle of continuous improvement [12] To measure attitudes and opinions, the most popular tool remains the survey. Public opinion polls and attitude surveys can be conducted and compared to benchmarks to determine whether the messages and behaviours of an organization have had the intended effect [52]. However, we must first understand the approaches to take.

## The right approach(es)

### Story telling

Story telling is by far and away one of the most important features of our everyday existence which taps into our innate ability to imagine and create mental images, engaging us through sequences of imagery that are not physically present, which is a fundamental aspect of human psychology [53]. Narrative transportation theory suggesting that storytelling is one of the best ways to engage with others, as it involves both emotion and cognition, leading to affective responses and enhanced engagement [54]. Stories help forge and enhance relationships exposing our common experiences and perspectives [55]. The application of storytelling can have different objectives such as creative (i.e. non-function) or to be persuasive (i.e. commercial Brand building). However, there is a body of literature on story telling as *engagement* understanding storytelling and engagement to be dialogic, and circular rather than linear. Leading to enhanced representation and engagement in domains such as policy engagement [56], education [55] and healthcare[14].

In the technology and science domain real stories of those working in specific elements of an industry gives rise to trust building. This has been found in responding to vaccine hesitancy in the Covind-19 pandemic by using local health care professionals [57], it has been found by increasing diversity in the professional research team in increasing diversity among participants in health research [58], and it has been found in the bringing together of the public and scientific professionals in STEM Cell research leading to findings that the public have the capacity to engage in sophisticated and nuanced discussions over science and technology [7].

In analysis of efforts of Parliamentary story telling Prior et al (2022) argue that engagement story telling falls into three steps of consideration:

1) Storyteller/ narrator
2) Characters and plot
3) The audience [56]

In the context of engagement, it is useful to understand the context in which stories are generated, by whom, and with what aim(s) .... *who* is the storyteller and narrator is important here [56]. What goes without saying is that those you are seeking to engage with must be part of the telling of the story, however there are nuances in how this is achieved. Such nuances can impact the overall outcome and the ability of the story to facilitate further 'engagement'. Societal contributions can narrate (through their own experiences) a story being told by others or the contribution of the public can determine the way in which the story is told. Put simply the storyteller controls the plot (in the form of sequence and shaping principles) [56] and this position of influence should be recognised in order to build stories that facilitate engagement.

Moving to character/ plot as identified by Christopher Booker (2004) there are seven universal patters in how story plots are told of: Overcoming the Monster, Rags to Riches, The Quest, Voyage and Return, Comedy, Tragedy, Rebirth. In the context of engagement Miller-Day and Hecht (2013) propose a narrative engagement theory for developing engaging messages around healthcare. Within this they refine the idea of plot and character in the engagement context to 'plot realism' (the plot must be realistic and believable to engage audiences, providing authentic models for vicarious learning) and 'character identification' (audiences should feel similar to or identify with characters, fostering cognitive and emotional empathy). This leads to interest, and a combination of interest, realism, and identification leads to engagement, which is crucial for the narrative's impact [14].

Finally the concept of 'audience' remains invaluable to the study of engagement as it moves us from considering communication as 'dissemination' to one of encouraging dialogue, which if successful moves the audience into the story itself and builds a dynamic between the two. This is achieved in a myriad of ways. Stories should be directed at the individual using language and sentiment such as 'you' and 'your' [56], however noting findings in the previous section this must be within the wider context of plots and characters that are realistic and identifiable. Further recommendations in the digital age are to create 'story worlds' where multiple co-creators develop stories in the digital space where the story emerges to be far more powerful than one way communication, building on a collective of emotional and cognitive engagement and an emphasis on quality over quantity [54]

Its notable that relating traits of the story to the individual themselves is a cornerstone of engagement. Technology is complex, self-driving technology is complex. Reorientating the '*technology'* idea of self-driving technology to an idea that is relatable to the public requires *the public* themselves to be engaged. This does not imply that the core facts relating to self-driving technology should be altered more that is should be interpreted by the public. Domain experts can therefore play an important facilitating role in firstly developing baseline material from which the 'story' of self-driving technology can unfold in partnership with the public, and secondly, they can act as a gatekeeper of the binary facts relating to the technology as those stories unfold. Within these parameters its likely that utilising the story approach could generate valuable engagement material. We will now further explore mechanisms of stakeholder engagement and co-creation to enhance such a strategy.

## Stakeholder engagement and Co-creation

A stakeholder is any actor (group or individual) that is influenced by a decision and/ or that can influence on that decision [15]. It is therefore key that we understand why stakeholder engagement matters and how it can be undertaken effectively. Stakeholder engagement matters because of:

- Trust: engagement fosters trust with stakeholders and in the outcome
- Accountability and Responsibility: stakeholders ensure accountability, transparency, socially and ethically responsible actions.
- Relationships: Effective stakeholder engagement enhances relationships and enables accurate insights and influence.
- Legitimacy: Co-creation with stakeholders enhances legitimacy and the likelihood of successful outcomes [15], [46]

Kivits and Swang's (2021) extensive work on stakeholder engagement in the aviation industry takes a more specific perspective on this safety critical eco-system highlighting the role of stakeholder engagement in addressing dynamic relations and complex environments that result in different frames of reference, networks, specialist insights, and ongoing changes [15]. Together with a traditional stakeholder mapping exercise Kivits and Swang (2021) suggest a stage of analysis should be undertaken to consider networks within which a stakeholder exists, their relative power and influence so it can be calibrated for, and their frames of reference [15]. Here we see further support that all information in all forms is likely not the solution, but mechanisms for accessing information as required so the right information at the right time (in the right format) can be determined through trustworthy Behaviour.

Furthermore, support for balancing power-imbalances is found by Bates et al (2010) finding that programmes of trust building are more successful when intended efforts to 'level the social and cognitive authority' of experts are implemented [7]. The public have a deficit of knowledge, and that lack of knowledge can drive rejection [6]. As we observed in the second case study, nuclear power, effective engagement must account for these imbalances and make efforts to address them, often in difficult circumstances.

Such stakeholder analysis of the UK public in the mobility industry has been undertaken at a regional [59] and national level [60] through the lens of 'personas', providing a useful baseline for with self-driving technology stakeholder mapping. Furthermore, specific industry stakeholder mapping in the self-driving sector itself has been undertaken in the form of supply chain structure [61].

A reoccurring theme within the engagement literature is co-creation. Co-creation is characterised by the three Co's Framework of: Co-Define, Co-Design and Co-Refine. Those who take part in co-creation processes are recommended to be called co-creators, with less focus on "empowerment" and more about facilitating people to harness the power they already have [62]. We have seen these characteristics in the symmetrical model [50], in the need to challenge assumptions and bias in stakeholder engagement [7], [15], [46] and through the inclusion of the public in the creation and telling of stories [14], [54], [56].

A challenge in any engagement context is balancing expert knowledge and societal contribution in ways that provide effective context and direction but do not impose bias and actively and meaningfully bring the public into the engagement processes. When done well the public have the capacity to engage in sophisticated and nuanced discussions over science and technology [7]. Miller-Day and Hecht (2013) propose a 'narrative intervention processes' for developing engaging messages around healthcare [14] which has been adapted below:

- Collecting narratives
- Analysis of narratives
- Stakeholder advisor input
- Content development
- Booster content

This model creates mechanisms to ensure engagement material remains factual and grounded, but for the actual material itself to undergo a process of co-creation. A key factor of importance to recognise is that the inclusion of key stakeholders at the start and throughout these processes will lead to those stakeholders themselves getting behind the final content. Stakeholder will have a stake in the final result and actively 'booster' its distribution and further engagement.

## Other tactics

These approaches provide signposts that move us towards an operational phase of realising these engagement outcomes including, segmenting the audience, creating communication based on self-interest, and the selecting of communication channels [52]. The most creative element in the strategic planning stage is the tactic [52]. Tactics are the specific communication *tools* and *tasks* that are used to execute the strategy. For example, in a seatbelt engagement campaign, the tactics would be the elements found in the educational kit, such as crossword puzzles, colouring books, or interactive games [52]. To develop such tactics, we must develop a strong understanding of the users perspective which we will explore in the following phase of the paper.

# Enlisting the help of the public (section conclusion)

As at the end of the previous phase of the report it is necessary that we consolidate the findings so far with a focus on the engagement objectives of this paper, to focus the discussion moving forwards. What are the guidelines and best practice for generating engagement programmes and materials? Exploring the difference between 'good' and 'bad' engagement reflected many of the traits found in the prior trust discussion, creating a context to define the key 'approaches' in the creation of engagement programmes.

- Engagement strategies fall along the continuum between notions of engagement as control and engagement as collaboration. **Effective engagement must fall at the 'collaboration' end of this scale** as ineffective stakeholder engagement has proven to led to a fall in trust.
- Effective engagement is characterised in the opportunity for stakeholders to **actively contribute and affect outcomes** (not just listened to).
- **Research** makes engagement activities strategic by ensuring that activity is specifically targeted to societal groups who want, need, or care about the information.
- The symmetrical model undertakes research to increase its understanding, but it **doesn't seek to use this information to persuade. But as the base for the building of mutual understanding** between the public and organisations.
- The degree of engagement of a target public groups (from passive to active) determines how, at a given point, engagement should be undertaken. If well-constructed **the public will progressively become more active (engaged)**.
- Storytelling is one of the best ways to engage with others, as it involves both emotion and cognition. A body of literature on **story telling as *engagement*** understands storytelling to be dialogic, and circular rather than linear.
- Those you are seeking to engage with must be part of the telling of the story, which **should in part determine the way in which the story is told**.
- Stakeholder engagement addresses dynamic relations and complex environments that result in **different frames of reference, networks, specialist insights**, and ongoing changes.
- The public have a deficit of knowledge, and **that lack of knowledge can drive rejection** if not accounted for. Mechanisms that **level the social and cognitive authority of experts** leads to enhance engagement.
- Co-creation is characterised by the three Co's Framework of: Co-Define, Co-Design and Co-Refine. Those who take part in the co-creation processes are recommended to be called co-creators, with less focus on “empowerment” and more on **facilitating people to harness the power they already have**.

- **The most creative element in the engagement planning stage is the tactic**. Tactics are the specific communication *tools* and *tasks* that are used to execute engagement.

We have developed an applied understanding of how good engagement is characterised, leading to a range of approaches for executing on this definition. Many of the elements explored in this phase of the paper reflect those identified (at a slightly higher level of abstracting) in the trust phase of the paper. The central theme through both phases, that warrants further exploration, is the perspective of the individual we seek to engage with (i.e. ‘empowerment’, ‘building mutual understanding’, ‘relatable stories and characters’, and the ‘you’ in engagement material).  In response to these findings the third phase of this paper will move onto take a self-driving technology specific focus to explore what aspects of self-driving technology matter to the public.

# 4. Engaging on what matters to the public: self-driving technology

## Why we should understand what matters to the public?

Let us begin by contextualising 'what matters to the public'. Throughout this paper we have found evidence that the public perspective is very often different to that of domain experts, moreover their perspective can bring unforeseen value. Our discussion on trust found that the theoretical application of trust to self-driving technology specifically, quickly brought the user into focus leading to a necessity of regularly calibrating the public's knowledge and expectation of autonomous vehicles. We further build upon this in the engagement literature understanding that making self-driving technology 'relatable' to an individual's circumstance will enhance engagement. Whilst there is an imperative for domain experts to anchor engagement in the correct self-driving technology data, their social and cognitive authority should be levelled. Effective engagement therefore requires grounding in what matters to the public.

It's important to clarify that this grounding does not definitely define the outcome, but that understanding what matters to the public provides a framework upon which we can access and facilitate engagement. These areas should form a focus and should be addressed but also used as a platform to evolve the engagement into new areas. A key characteristic highlighted within the paper is that the public have a surprising capacity to digest complex information leading us to realise value from harnessing the power they already have. Nevertheless, the right information at the right time (in the right format) will be needed. Orienteering engagement around what matters to the public creates potential for ever more sophisticated conversations, greater trust, and moving the public into a progressively more active role of critique and advocacy. To increase the public knowledge of self-driving technology and move into this active role we must engage on what matters.

A mapping between the evolving nature of this approach and the dynamic nature of 'trustworthiness' found earlier in the paper should not go unnoticed. What matters to the public today is very likely different from what will matter to them in x years' time. This will be a dual characteristics of a) evolving prevalence of the technology leading to more knowledge among the public and b) evolving understanding between stakeholders, best achieved through engagement. As we learnt in the engagement phase of the paper, at any given time research should be undertaken to understand who wants, needs, or cares about the information and that research acts as a base for the building of mutual understanding. Within this phase of the paper, we therefore research the body of knowledge on public trust, Acceptance, and Adoption of self-driving technology to establish a baseline for engagement.

## Approaching the data: methodology

There have been extensive studies undertaken into understanding what is important to the public for the introduction of self-driving technology. The studies are in fact so frequent that search results on review papers of such studies return a very high frequency of papers. This leads us to the premise that these review papers, by amalgamating the findings of many different studies, will provide valuable insights. Whilst its likely these insights have some degree of generalisation it is important we begin from a robust base understanding. We further refine the objective of this phase of the paper to build a 'framework understanding of what matters to the public that will act as an actionable tool for engaging specific segments of the public. We will address this challenge by drawing findings from an amalgamation of many studies (review papers) and then unpacking in more details the factors that this process identifies.

Key word searches for 'Review' or 'Summary' or 'Status' or 'Systematic' or 'Comparison' and 'trust' or 'Acceptance' or 'Adoption' and 'Automated Vehicle' or 'Autonomous Vehicle' or 'Self Driving Vehicle' were undertaken. Given review papers are an amalgamation of research that has come before them review papers for 2023 and 2024 were given priority (although other papers, especially those with a novel focus were also given consideration).

### *Quantifying what matters to the public*

It's not just the volume of papers that presents a challenge, but also the range of factors found to be important to users. There is a noticeable split in focus between general self-driving technology generally, private self-driving technology and shared/ public self-driving technology with further sub categories of robo-taxies, robo-shuttles, robo-buses and pooled robo-taxies [14], each leading to slightly different outcomes. Two examples of these wide ranging findings are firstly Bala et al (2023) finding that public acceptance falls into four classes: 1) demographic factors (e.g. gender, socio-economic) 2) level-of-service attributes (e.g. ticket/ride fare, access distance and waiting time) (3) travel behaviour patterns (e.g. commute type, frequency of private car use). Secondly Greisfenstine et all (2024) published a table of all 63 factors they found in the literature shown in Figure 3: Example finding of <u>all</u> factors influencing acceptance, trust, and adoption of self-driving technology (taken from Greifenstein et al 2024) [63].

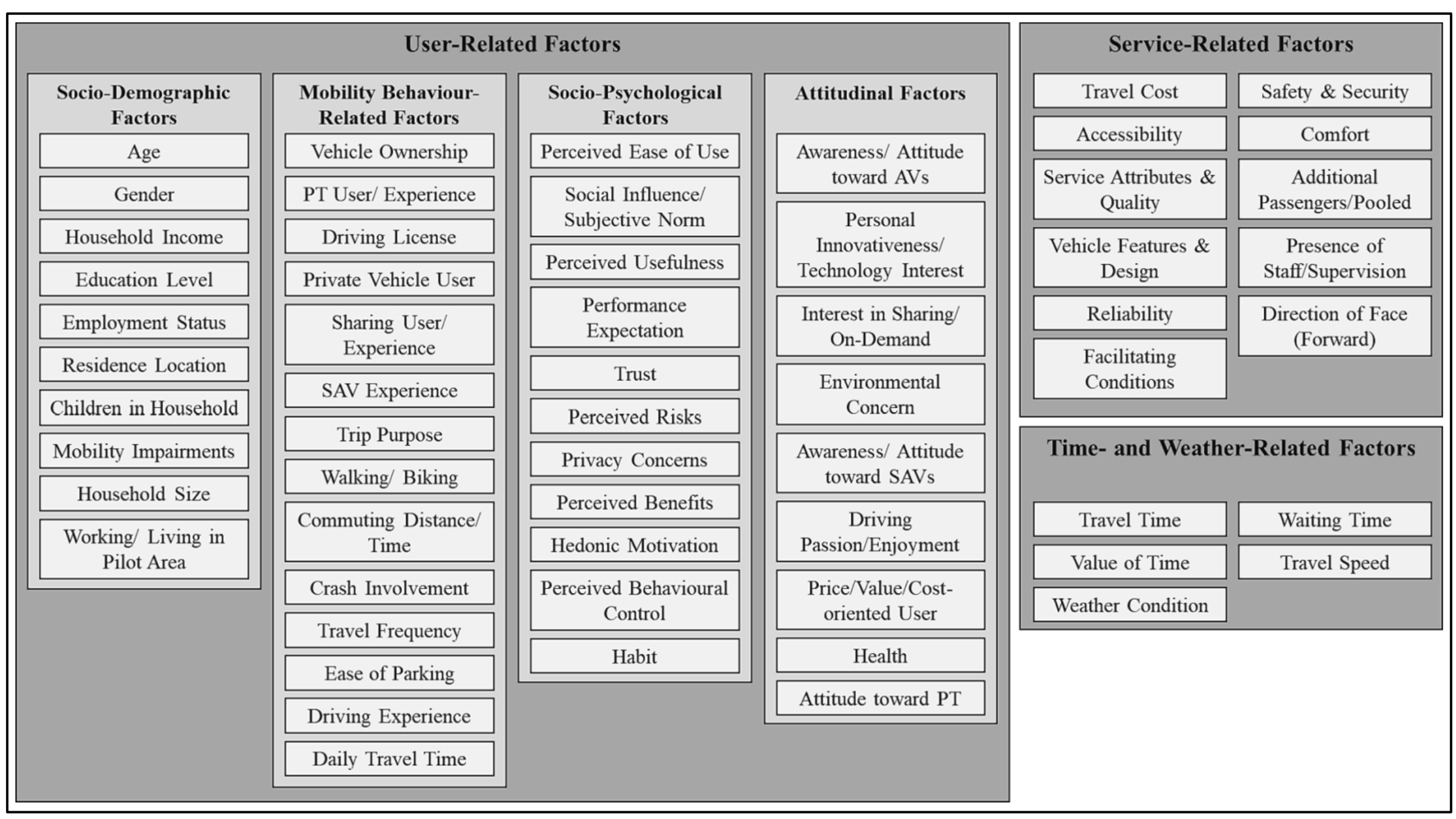

*Figure 3: Example finding of all factors influencing acceptance, trust, and adoption of self-driving technology (taken from Greifenstein et al 2024)*

Given the scale of data, a number of studies recognise the value of quantifying the factors of importance in self-driving technology to the public. This approach builds on the meta-data made available from studies and statistical analysis. In the context of this review such quantifiable conclusions would enable us to prioritise and focus in on a sub-section of actors that are identified as the most important and relatable to the public. Such an insight would begin to build a scalable framework around which key information and engagement strategies could be built.

Rahman et al (2023) review the meta-data of 81 literature sources [64]. The review highlights that both internal and external influences are important, where internal factors relate to the individual (such as attitude, experience, travel behaviour) and external factors relate to the wider environment (such as infrastructure, incentives, congestion). Rahman et al find that 'opportunities and challenges' become the central framework of users perspectives, with a mix of internal and external factors contributing to the overall construct of those opportunities and challenges.

Mansoori et al (2023) also undertake a comprehensive systematic meta-data review of factors influencing the adoption of Autonomous Vehicles (AVs) [65]. The review meticulously analyses 71 studies focusing on various frameworks used to investigate self-driving technology adoption. The review categorises the factors into nine distinct categories of psychological, technological, social, environmental, security, AV-related, risky, conditional, and monetary factors, providing a taxonomy for future research. Quantifying these outcomes using meta-data analysis the study proposes an adoption framework that aims to offer a holistic depiction of the variables and hypotheses examined in previous self-driving technology adoption studies, thereby providing an overarching view of this subject matter. This model is show in Figure 4 Analysis of Mansoori et al (2023) proposed SDV vehicle adoption model (blue dashed lines and blue text added by the authors).

This model clearly shows that the most studied target variable is 'Behavioural Intention'. 'Intention' as this is a result of most studies not exposing participants to actual self-driving technology experience and highlights the need for future studies to validate those factors influencing behavioural intention flow into actual use. However, this does provide some early indication of how factors may change over time as we move from most users being in a pre-adoption stage (intention) to a stage of adoption and usage (actual use, willingness to use, and willingness to pay). These shifts will require further research as technology availability increases.

Whilst a wide number of factors are identified among the literature four far exceed the others on their influence of Behavioural Intention which we go onto understand in more detail.

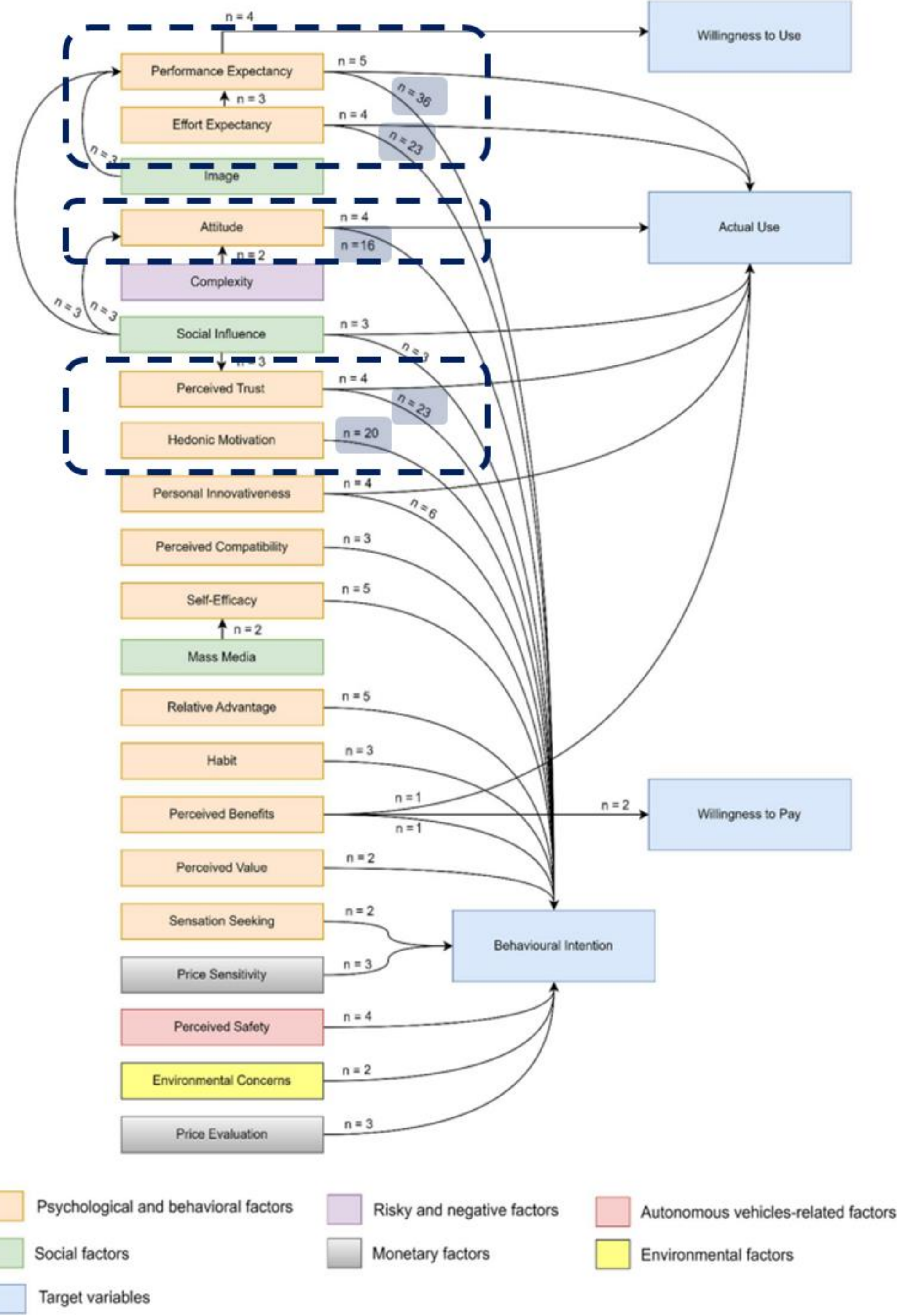

*Figure 4 Analysis of Mansoori et al (2023) proposed SDV vehicle adoption model (blue dashed lines and blue text added by the authors).*

## What matters to the public?

### How does self-driving technology benefit me?

Performance expectancy is defined as the degree to which an individual believes that using the system will help them to attain gains (professionally or personally) [66]. We can derive more specific self-driving technology insight on this definition from the questions used in studies. From studies that published their questionnaires there is a focus on questions such as getting passengers to their destination safely, comfortably and punctually, improvements in quality of life i.e. improving social and family life, productivity, and improvements to wider society such as reducing congestion, reducing emissions, and improving the transport system [67], [68], [69], [70], [71], [72], [73].

Studies differ in their interpretation of Performance Expectancy as individual benefits i.e. getting you to your destination comfortably or as a external benefit i.e. the influence on congestion. Understanding the roles of each in influencing behavioural intention will be a useful understanding in forming engagement programmes. It is worth noting both perspectives have alignment with the 'business case' often discussed in other aspects of the self-driving ecosystem. Articulating a clear need for self-driving technology and the 'pull' factors is an imperative not only for the public, but also policy makers and investors.

The way in which such benefits interact with other factors is complex. Once consumers determine that self-driving technology is a trustworthy mode of transportation whose advantages outweigh its disadvantages, consumers will consider whether this advantage is suitable for their own needs [67]. We should however be conscious that the concept of self-driving technology is likely insufficiently mature for the public to make an accurate judgement on its benefit to them. In this regard we should be mindful of 'distinguishing the relative hype from the reality of experience' as the construct of Performance Expectancy is sufficiently broad to be interpretated inconsistently leading to a need for a better understanding of this construct as applied to self-driving technology. [74].

By way of comparison a study with a focus on logistics found high levels of clarity among participants in specific areas of performance expectancy, for example addressing the specific challenge of reaching areas which are difficult to access by vans or cars [75]. It is likely clearer to a logistics professional how a self-driving technology would help within their daily work as opposed to the often more diverse scenarios of public or private transportation. This illustrates the need to enhance specific mobility-of-people use case and evidence behind the performance claims.

Whilst performance expectancy has been little studies specifically, Sweet et al (2022) use this lens to consider intentions to use a private self-driving technology verses intention to use a self-driving technology shuttle service. Performance Expectancy of a Private self-driving technology was found to be most appealing to those from 'households with higher incomes and with more household members .. that have complex travel behaviours and fewer members with driving licenses' [74]. For this group previous experience with self-driving technology related technologies i.e. ADAS, was found to be influential in the willingness to adopt. However, on the other hand those with prior transit experience (i.e. public transport) defined as holding a travel pass, were found to perceive most performance expectancy benefits from a self-driving technology shuttle service regardless of other demographic characteristics [74]. Prior experience of a self-driving technology shuttle did not enhance their willingness to adopt further [74] illustrating a difference in prior self-driving technology experience on willingness to adopt across private and shuttle use cases.

In the context of engagement understanding performance expectancy helps us understand how the public are likely to relate to the technology. Developing material that addresses and opens up questions for discussion such as how would self-driving technology get passengers to their destination safely, comfortably and punctually, how self-driving technology would deliver improvements in quality of life i.e. improving social and family life, productivity, and how self-driving technology will deliver improvements to wider society such as reducing congestion, reducing emissions, and improving the transport system [67], [68], [69], [70], [71], [72], [73]. However, in communicating such issues with wider society its key we 'distinguishing the relative hype from the reality of experience' [74]. Many of the perceived benefits envisioned from self-driving technology lack evidence to support the claims. Given that Performance Expectancy is the single most influential predictor of behaviour careful consideration to how these benefits are communicated relative to the evidence is warranted, together with a concerted efforts to develop such evidence as the technology evolves.

## How will I use a self-driving technology?

Effort expectancy is defined as the degree of ease associated with use of the system [66]. From studies that published their questionnaires Effort Expectancy is considered through questions such as mental effort/ ease of comprehension, comfort, compatibility with other technology, learning time to become skilful at using it, ease of payment, ease of access [67], [68], [69], [70], [71], [72], [73].

These studies have taken different approaches to exposing their participants to self-driving technology. The most advanced studies were aligned with technology demonstrations and participants took a ride in self-driving technology. However, given the phase of development we are in even that was a luxury, and most studies only conveyed the conceptual idea of a self-driving future. This makes a true understanding of effort expectancy hard to understand. One interesting finding was that effort expectancy in a self-driving bus scenario was drastically different with and without a conductor on the bus. The fact that the bus is operating on its own is much more evident when passengers step into vehicles without any employees on board [70]. This relationship between support and effort expectancy is at odds with the idea of autonomy (i.e. no driver). This has important implications for our understanding of trust and illustrates that expectancy and trust are related [67].

In terms of engagement it is expected users will have questions regarding the degree of ease in using the system, which will influence their intention to use self-driving technology. Furthermore, this effort expectancy has a relationship to overall trust in the system.

## How will interacting with self-driving technology feel?

Hedonic motivation is defined as the fun or pleasure derived from using a technology [76]. From studies that published their questionnaires Hedonic Motivation was considered through questions such as it would be fun and enjoyable to ride in a self-driving technology, using a self-driving technology is exciting, to be driven by a self-driving technology would be pleasant [69], [73], [77]. There is a notable range here from concepts of 'fun' to concepts of 'comfort' which would benefit from further understanding through the research. In this regard its worth noting that specific literature on hedonic motivation demonstrates a bias towards younger people (samples with most or all participants under 35 years of age) [69].

Interestingly the literature on technology 'adoption' undertook a shift from the professional environment to the consumer environment in response to the raise of consumer goods such as mobile phones and personal computers in the 2000's. It is with this shift hedonic motivation came into play (having not been shown to be a primary factor of influence in a professional environment).

One study in this area found that around a third of respondents believed a multimedia offering and internet access would make the trip more enjoyable [69]. However, this multimedia and internet offering illustrates there is a grey area between the self-driving technology technology itself and wider technologies it takes advantage of. This suggests the public do not conceive them to be different. Furthermore, within hedonic motivation itself there is a difference between in vehicle technology that enables the public to be more productive i.e. undertake professional work, and to be entertained i.e. watch a film.

The limited specific literature finds social influence and perceived safety to have the strongest links with hedonic motivation [78]. Social influence here is defined as the extent to which consumers perceive that important others in their life believe they should use a particular technology. Theories linking social influence and hedonic motivation argue that these social influences directly and indirectly form mental models and expectations that the individual seeks confirmation of and that enjoyment is a social construct in itself where by enjoyment in a particular activity becomes a shared social experience [78].

An additional study found hedonic motivation to have a higher correlation with private self-driving technology ownership (as opposed to public self-driving technologys which showed a higher correlation with utilitarian outcomes) [79].

Hedonic motivation is considered in the literature as a motivator in which users actively seek pleasure from self-driving technology especially relevant for the younger generation below 35 [69] and private self-driving technology [79]. However hedonic motivation has also been found to be an indirect barrier to self-driving technology whereby discomfort creates a barrier to other positive factors such as social influence or personal security [78]. Such factors can be perceived and/ or experienced. This perspective of comfort (perceived and experienced) provides insight to the types and forms of content to address in engagement material.

## Your balance of favourable and unfavourable evaluations of self-driving technology

Attitude is a complex term and has been defined as "the degree to which a person has a favourable or unfavourable evaluation or appraisal of the behaviour in question" [66]. However, these factors can play out differently in both a pre and post experience scenario. For example, 'attitude and intentions can be triggered by objects and clues in the environment' [76] suggesting that experience plays a role in attitude formation for new adoption decisions. Further emphasised by influencing habit formation [76].

As noted by Mansoori et al (2023) a number of self-driving technology 'studies provide evidence that attitudes are affected by various variables, including social influence, hedonic motivation, performance expectancy, habit, price, trust, experience satisfaction, and effort expectancy' [65]. Self-driving technology attitude studies have found that 'there are some users that are entirely hostile to the prospect of AV's' [80], and a 'large fraction of the population is not yet ready to use [self-driving technology] with no driver' [81]. More encouragingly attitude studies with real-self-driving technology trials showed users with less favourable views improved after having experienced a self-driving technology [82], [83], [84], [85].

Attitude studies have also shone a light on moral and ethical dimensions of self-driving technology (which is also closely aligned with trust). These studies considered the impact on users' intention to use a self-driving technology based on selfish collision decision making algorithms (prioritise the passengers safety) verses utilitarian collision decision making algorithms (saving more lives over fewer, regardless of whether they are passengers or pedestrians) [86]. Finding that the general acceptance of self-driving technology aligned with a preference for utilitarian decision making, but specific use of a self-driving technology by an individual was associated with selfish decision making of the self-driving technology [86], [87]. One of these studies used the example of a self-driving technology-police car as the utilitarian example [87], this demonstrates a shift in perception between private and public operated vehicles also. In this regard services such as self-driving technology buses, shuttles, and taxis can be shown to warrant further understanding. Understanding how moral and ethical issues are interpreted by the public will be important to the forming of engagement programmes.

Attitude is commonly associated with past experience and habit, two factors that are closely tied to an individual's specific mobility experience and needs. A range of 'attitude' studies usefully focus in on specific scenarios including small urban areas showing the importance of location. Specific studies finding that 'technology-savvy respondents were more positive but had more concerns about [self-driving technology]-transit integration than others. Respondents who enjoyed driving were not necessarily against [self-driving technology]', [88] and that low-income households were associated with a 25% increase in the likelihood of using self-driving technology in colder climates where noting having a car or using public transport more frequently create a greater personal challenge [89]. Others focus on differences across age groups finding that 'there is a generational difference in mobility attitudes between millennials and Generation X.  where preferences for transit and alternative modes and less reliance on private vehicles among Millennials were more a reflection of their preferences in lifestyle choices and not so much constrained by their socioeconomic status' [90] and that seniors (65 years and older) cluster into three distinctive groups of those with 'a negative perception towards AV's', 'those with a positive perception towards using shared AV's, but a negative perception of AV-pedestrian interaction', and 'those with a negative perception towards shared AV's, but a positive perception towards pedestrian-AV interactions' [91].

Thinking how these findings on 'Attitude' could underpin engagement programmes is intuitively less clear. Attitude appears to be a broad concept that manifests based on specific factors such as mobility type, demographic, and experience. Attitude can therefore be conceived as a theoretical underpinning that has many similarities to trust i.e. a balance of positive and negative factors. Indeed, we found alignment with moral and ethical questions, and with the fact attitudes are far from always positive (drawing out our prior findings of embracing sites of contention). Attitude is also influenced by past experience and habit, often modified by demographic factors such as age.

## Do I trust self-driving technology?

Perceived trust is the fifth and final attribute to standout above all others as important to users. This neatly brings us full circle from the extensive discussion on trust at the beginning of this paper. Given the first phase of this paper spent some time discussing trust in the specific self-driving technology context we will not revisit it here.

## Limitations in the data

A universal challenge that all studies recognise is that self-driving technology are not available for regular experience by the public. Most studies are found to be survey based and with information describing what a self-driving technology is and would be like. This lack of realism in text heavy surveys leads to potential weakness and unreliability [92]. Furthermore, those studies that do include exposure to self-driving technology do so in a controlled environment often with a slow-moving vehicle. Few studies investigate possible bias or miss-understanding on account of having not experienced the technology [93]. Lack of actual experience leads to a range of unanswered research questions including how new features and technologies influence user intention and product design, how prior experience of partially automated systems (i.e. ADAS) influence trust, social experience, and how challenges around data, security and safety experience influence perceptions [65].

Other notable gaps in the literature are the perspective of pedestrians, cyclists and other vulnerable road users (VRU's). For example these road users are especially reliant on informal gestures as part of their interactions with other road users and more should be understood about how these factors will be replaced with the introduction of self-driving technology [94]. VRU's perceptions of self-driving technology needs to be better understood [93], [94]. In addition there is still a need for better understanding the effects of demographic characteristics—e.g., income, education and age—for example self-driving technologys as first mile/last mile solutions are more likely to target transport-disadvantaged populations such as senior citizens and people with disabilities [95] research that best identifies and targets early adopters, that understand the diversity of preferences across socioeconomic status, will help developers, governments and stakeholders segment, target, and promote self-driving technology [64]. This brings into focus ethical, legal and insurance implications of intention to use self-driving technology which is also little explored [93], [96] such as the idea that self-driving technology will be travelling together with driven vehicles [93]. More needs to be done to understand societal interpretation and understanding of the ethics of self-driving technology [94].

In response to a lack of lived experience of self-driving technology and to promote a positive acceptance of them before they are available strategies to build awareness are suggested such as via public media campaigns and hands-on tests [97]. However few studies discuss how to communicate better with respondents such as approaches to introduce basic functions and application scenarios [93].

These factors will also change over time as self-driving technology become increasingly prevalent in public life. It is of great necessity to include a spatio-temporal perspective through doing regular surveys which will examine how people's attitudes towards and demand for new mobility technologies and services change over time [98] and one recommendation is to research survey the same panel of respondents repeatedly over time [64], [92]. In addition, much more work need to be done to understand differences across VRU's [94] [95], specific self-driving technology types and services [93], [96], and over time as the technology improves [98].

## Towards a framework of understanding what matters to the public (section conclusion)

It is necessary that we consolidate the findings so far with a focus on the engagement objectives of this paper. Routed in findings from the previous phases of the paper on trust and Engagement we have identified that the public perspective on self-driving technology is very often different to that of domain experts. What self-driving technology element matter to the public? Starting with a review of research studies into self-driving technology trust, Acceptance, and Adoption we have been led to answers:

- What matters to the public today is very likely different from what will matter to them in x years' time
- **Orienteering engagement around what matters to the public today**, creates potential for evolving ever more sophisticated conversations, greater trust, and moving the public into a progressively more active role of critique and advocacy
- **Extensive studies have been undertaken** to understand what is important to the public for the introduction of self-driving technology. In some instances, **over 63 factors** of importance have been identified. This paper focuses on review papers that **statistically evaluate meta data** of over 70 studies to quantify the factors of importance to users
    - Whilst its likely these insights have some degree of generalisation it is important we **begin from a robust base data-driven understanding**.
- We find five factors of important consistently shown to matter to users:
    - **How does self-driving technology benefit me?** (known in the academic literature as Performance Expectancy) indicating a need to engage the public in questions such as:
        - How would self-driving technology get passengers to their destination safely, comfortably and punctually
        - How self-driving technology would deliver improvements in quality of life i.e. improving social and family life, productivity
        - How self-driving technology will deliver improvements to wider society such as reducing congestion, reducing emissions, and improving the transport system
        - Highlighting the need to **distinguish relative hype from reality** and clearly articulate where responses can be justified with evidence and where they cannot. **Unjustified claims have been shown to lead to mistrust.**
    - **How will I use a self-driving technology?** (known in the academic literature as Effort Expectancy) indicating a need to engage the public in questions such as
        - What will be the mental effort/ ease of comprehension?
        - How comfortable will I be?
        - What is the compatibility with other technology?
        - What is the learning time to become skilful at using it?
        - How easy is it to access and pay for?
        - One interesting finding was that effort expectancy in a self-driving bus scenario was **drastically different with and without a conductor on the bus**. The fact that the bus is operating on its own is much more

evident when passengers step into vehicles without any employees on board

- **How will it feel to interact with a self-driving technology?** (known in the academic literature as Hedonic Motivation) indicating a need to engage the public in questions such as
    - Would it be fun and enjoyable to ride in a self-driving technology
    - Will using a self-driving technology be exciting,
    - Would to be pleasant to be driven by a self-driving technology
  - Theories linking social influence and hedonic motivation argue that **enjoyment is a social construct**
  - Perceived safety is related to 'positive outcomes' whereby **fear can impede the spontaneity and freedom** associated with hedonic motivation
- **Your balance of favourable and unfavourable evaluations of self-driving technology** (known in the academic literature as Attitude)
  - Attitude is influenced by past experience and habit, often modified by demographic factors such as age.
  - self-driving technology attitude studies have found that '**there are some users that are entirely hostile to the prospect of AV's'**, and a 'large fraction of the population is not yet ready to use an [self-driving technology] with no driver'. More **encouragingly attitude studies with real-self-driving technology trials showed users with less favourable views improved after having experienced a self-driving technology**
  - General acceptance of self-driving technology aligned with a preference for utilitarian decision making, but specific use of a self-driving technology by an individual was associated with selfish decision making of the self-driving technology. One of these studies used the example of a self-driving technology-police car as the utilitarian example, this **demonstrates a shift in perception between private and public operated vehicles also**

In the context of engagement this phase of the paper has found evidence for the foundational issues of importance. This does not in any way preclude the other factors, however it demonstrates that these are the foundational factors upon which other nuanced and more specific factors relating to an individual and/ or use case can then be built upon. This focus on foundational factors is key to creating a scalable framework of engagement. Having scaled from a theoretical understanding of trust, through the application of Engagement and finally the specific analysis of what about self-driving technology matters to the public we must now bring it all together. The final phase of this paper will draw on all the threads of learning to propose a framework for the creation and implementation of a self-driving technology engagement programme.

# 5. A playbook for self-driving technology education, engagement and trust (discussion, recommendations and conclusion)

What have we learned and how can it be applied? We set out at the beginning of this paper to identify how Government and industry can prepare for wide scale awareness of self-driving technology among the public. To bring the public on every step of the journey a theoretical grounding in trust has been established, leading to 'education and engagement' programmes as an effective mechanism through which to establish the public role in the building of a trustworthy system. Robustness is built into this approach through understanding what matters to the public and acknowledging this is often different to what domain experts consider important. A bridge between the public and domain experts is built by beginning engagements on what matters to the public which creates a platform for new depths of content to materials. At each stage domain experts should acknowledge and calibrate for their relative knowledge and power advantage to sustain effective engagement throughout this journey.

Its necessary that we arrive at a conclusion within this paper that is implementable and scalable by practitioners. This final phase of the paper will therefore translate the findings throughout into a model for the implementation of self-driving technology education and engagement programmes. This aligns with the 'narrative' review objective of this paper to develop a model [6] of inclusive and accessible self-driving technology engagement. Drawing on threads of our learning this will begin with an overall framework for engagement and trust, becoming progressively more concrete in its application by systematically organising the attributes that form the public engagement journey with self-driving technology, together with a framework for co-creation, finally arriving at a model of scalable self-driving technology engagement across the stakeholder landscape.  Let's begin with the overall Framework.

## A framework for engagement and trust

A shift from notions of trust to trustworthiness clearly emerged through the first phase of this paper, characterised as trustworthy behaviour. Through case studies in trust and an exploration of engagement in the second phase of the paper we were able to define attributes of trustworthy engagements. These attributes form a set of overarching principals which are:

- **Transparency, reliability, explainability, and representative diversity**: Technology 'must advance in a way that meets and exceeds existing regulatory frameworks and societal expectations' [34] by adopting fundamental characteristics of transparency, openness and accountability [6]. The role of explainability – the self-driving technology ability to describe the rationale behind their outputs in human-understandable terms had the most profound impact on trust [43].
- **Engagement is the enabling of a meaningful response**: (through cognitive involvement and community capacity building [36] wherein the public and organizations who share mutual interests in salient topics … aim at goal attainment, adjustment, and adaptation for both their benefits) [8].
- **Levelling of biases/ power/ knowledge deficits:** Power and influence are unavoidable between stakeholders and attempts to level the playing field support the building of trust [7], there is an 'overriding need for fairness in any consultation process' [37]. For example, in the Nuclear Energy Policy example policy makers elevated the importances of some elements they had not expected to be part of the policy framework until they heard first hand strong feelings expressed in the consultation [37].

- **Co-creation**: Co-creation is characterised by the three Co's Framework : Co-Define, Co-Design and Co-Refine. Co-creation facilitates people to harness the power they already have [62].
- **Do not try and promote trust and acceptance, promote engagement with its own outcomes**: trust is not absolute but a critical mass of negative and positive consideration [13]. Stakeholders must adhere to explicit and transparent principles of good practice [6] so individuals can arrive at their own level of trust judgement. If engagement programmes are well-constructed the public will progressively become more active, acting as informed critiques and advocates when highly engaged [8]. There is a notable desire of stakeholders to be involved in more dialogic forms of engagement [47].

| Principles | Tactics | Tools |
|---|---|---|
| **Co-creation**<br><br>**Representative diversity transparency, reliability, explainability**<br><br>**Engagement is the enabling of a meaningful response**<br><br>**Levelling of biases/ power/ knowledge deficit**<br><br>**Do not try and promote trust and Acceptance, promote engagement with its own outcomes** | Imaginaries of the future<br><br>Sites of contention as sites of engagement<br><br>Simplicity (which is not dumping it down)<br><br>Research (contextual and engaged)<br><br>Remove jargon<br><br>Use of 'direct' generated content to encourage 'wide' and 'amplified' response and engagement<br><br>Repeatable and tested core narratives/ messages<br>Distil the 'essence of it'<br>Understand the audience<br><br>Invite responses through incomplete material | Data/ evidence<br>Benchmarks (i.e. other industries)<br><br>Real use cases<br><br>Story telling<br><br>Story worlds<br><br>Situational analysis statements<br><br>Crowd sourcing<br><br>Audience self-interest/ personas<br><br>Relate to past experience (other technology/ travel patterns)<br><br>Address all learning styles : auditory, visual, and kinaesthetic<br><br>Stakeholder mapping<br><br>Artistic representations |

*Figure 5: A framework for engagement and trust*

To bring these principals into focus for practitioners we further evaluated the literature on engagement. The engagement literature explored 'tactics' and 'tools' that can be used to implement these overarching principals. In some instances, these approaches were specific to engagement i.e. 'imaginaries of the future' which give the public the opportunity to conceptualise how future self-driving technology products and service may manifest providing an input to the design cycle [6]. In other instances, these were well established methodologies in their own right that had been refined within the context of trust building and engagement i.e. 'Story telling' whereby those you are seeking to engage with must be part of the creating and telling of the story [56]. The overarching principals and a full list of Tactics and Tools are listed in Figure 5: A framework for engagement and trust.

Recommendations: This framework should be used as a reference for all stages of engagement and for establishing the public role in the building of a trustworthy system. This may be proactive (such as content design, workshop design, events, user consultations, technology trials) or reactive (addressing concerns, navigating conflict, challenging misconceptions, responding to media articles). It is not claimed that this is an exhaustive list, however it will act as an effective tool to plan and implement activities of engagement and trust building. It is recommended this framework be continually challenged and updated through primary and secondary research. To further aid practitioners we propose a specific self-driving technology model that will sit alongside this overarching approach. We turn to this self-driving technology framework next.

## Calibrating the public's trust on their journey towards self-driving technology

Let's combine this framework with our learnings from 'Engaging on what matters to the public: self-driving technology' to move into the realm of self-driving technology itself. It's important to emphasise that this is not a case of the public defining independently what matters but structuring the start of engagement around the public's perspective, [14] providing a platform for deepening the discussion between the public and domain experts. In the next stage of this discussion we will propose a model for self-Governance to implement these engagement principals at a concrete level, but let's first define the attributes for engagement that matter to the public as they calibrate trust on their journey towards self-driving technology.

Earlier in the paper we found that applying a theoretical understanding of trust to self-driving technology themselves quickly brought the user into focus [40] [4]. This focus broke self-driving technology trust down into two, trust *in* and trust *with*. To re-cap trust *in* the system, means the users trust in the capabilities of the system and/or in the system's ability to do what it is supposed to do. Here we are concerned with the formation of knowledge <u>before</u> using or being exposed to the actual technology. Trust *with* the system, means users awareness or attitude towards the system's limitations and their subsequent ability to adapt their <u>use</u> of the system to accommodate whilst still deriving the expected benefits [4].

Self-driving technologies are an emerging technology and we are on a journey towards them. At the level of each individual they are also on their own journey towards first contact with the concept of self-driving technology through knowledge acquisition, maturing to actual interaction or use over time. Reflected in the fact that trust is dynamic because of continued knowledge acquisition (and often old knowledge forgotten [11]) a model of self-driving technology knowledge acquisition would be useful. Such a model would provide guidance on knowledge attributes which engagement programmes could use to effectively hook into mental models as they form for those who want, need, or care about the information [12] Public openness is found to be a key pillar of trust [6]and as increasingly large volumes of information are made transparently available, we have concluded that engagement programmes must develop approaches to defining the right information at the right time (in the right format). Finally, our model must reflect the findings that the public have a surprising capacity to digest complex information and make nuanced judgments [7].

Figure 6: Calibrating the public's trust on their journey towards self-driving technology is built upon these overarching findings, and a range of specific findings throughout the paper. Having now justified the overarching motivation for this model from learning throughout, we will now discuss the specific learnings that underpin each building block of the model.

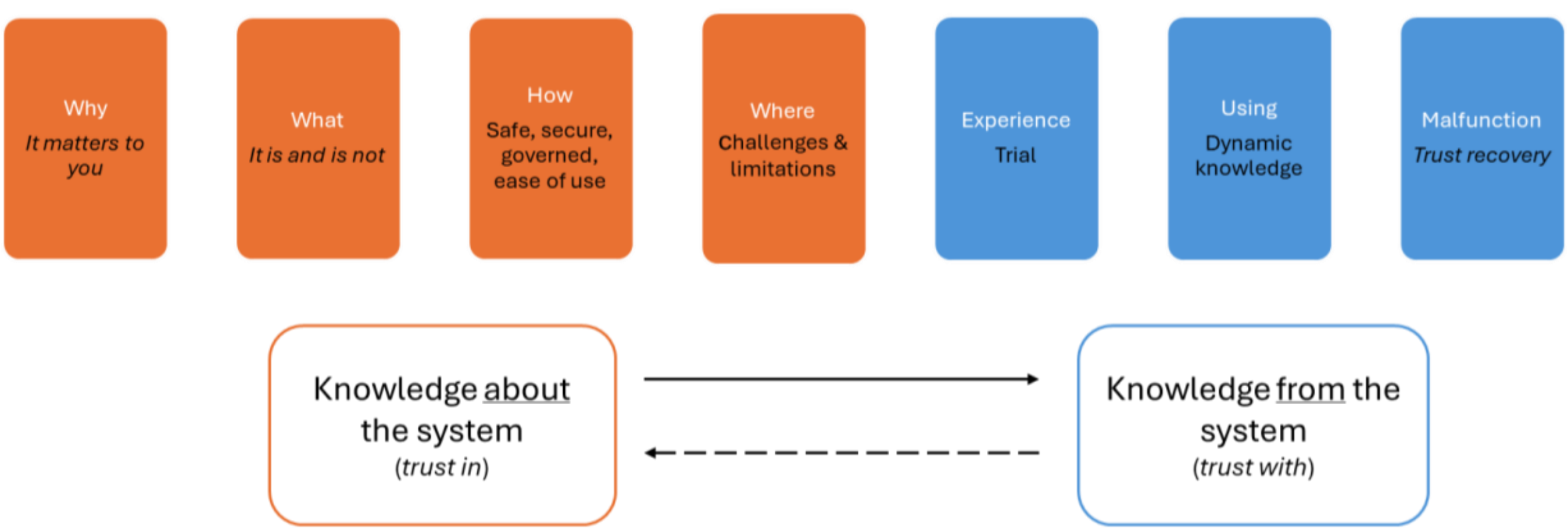

*Figure 6: Calibrating the public's trust on their journey towards self-driving technology*

**Knowledge about/ Trust *in***

- **Why: Benefits**: the most significance factor influencing intention to use self-driving technology is personal benefit referred to in the literature as 'Performance Expectancy' [65]. The move to intended actual use of a technology is heavily influenced by the personal benefits an individual derives from it [66]. Within the context of transparency, a pivotal characteristic of trust, its essential *perceived benefits* are grounded. For example, no commercially viable self-driving technology service or product is currently available in the UK. As a result, very little (if any, possibly some in simulated traffic modelling) evidence can back up any *perceived benefit* claim. A trustworthy programme of engagement should clearly acknowledge this status of the technology development cycle. This is important learning that brings together elements of all three phases of the paper and informs how to engage on 'benefits'. We can actually flip this challenge to become an opportunity drawing on engagement tools such as 'imaginaries of the future'. Personal benefit can be more easily conceptualised when the individual feels comfortable about their personal safety, security, and use of self-driving technology [78], dictating a natural progression to the subsequent components below.
- **What is a self-driving technology?** Which addresses the knowledge deficit between the public and domain expertise [6], recognising that at this time most of the public have little-to-no knowledge of self-driving technology [92], grounding the starting point for the self-driving technology engagement in the relevant stage of the public mental model development [14].
- **How safe, secure and easy to use is a self-driving technology?** Reflecting on the interwoven nature of trust and (personal) safety and security [4], we can build upon the finding that first consumers will seek to determine that self-driving technology is a trustworthy mode of transportation whose advantages outweigh its disadvantages before considering whether this advantage is suitable for their own need [67]. The literature has illustrated that these considerations can be complex i.e. ethical evaluations of personal vs utilitarian benefits [79], and comfort enhancing when a conductor is present on a Self-driving bus [70]. Finally ease of use referred to as 'Effort Expectancy' in the literature is the second most important determining intention to use self-driving technology. We can conceptualise this second component as an extension of the first 'What is a self-driving technology' recognising that once an individual has an abstract conception of a self-driving technology their attention shifts to personal impacts.

- **How: Governance and ethics**: Through the first section of this paper we found that trustworthy behaviour and trustworthy processes (along with the outcome) are central to the building of a trustworthy system [6]. This takes us beyond the technology itself to the wider systems-of-systems. The implementation of this theoretical lens manifests in forms of Ethics and Governance (informal and formal) [38]. One study indicated a contradiction in the public attitudes when they are in the self-driving technology (a preference for the safety of the passenger) and for the general Governance of self-driving technology (a preference for the least harmful outcome across passengers, pedestrians and other road users) [86].( Such contradictions indicate that ethics, standards and regulation are likely to be a highly contested area of debate among the public, yet in alignment with findings earlier in the paper embracing sites of contention is a critical step in the enabling of trust [37] [6].
- **When/ where: Challenges and limitations**: Echoing the lessons on perceived benefits in the prior bullet point, this step in the model goes further to acknowledge findings in engagement, where embracing sites of contention (when done correctly) enhances engagement [6]. Such challenges and limitations form the sharp end of the wedge for the public and once fundamental concerns over safety have been addressed, they very often become practical in nature. For example, where a domain expert may recognise the technical verification and validation of automated systems a challenge, the public are more likely to consider operational issues such as service frequency, cost, ease of payment and drop-off/ pick up locations as factors that ultimately determine acceptance [63]. Many of these factors are as yet indetermined and again being transparent means openness about this with the public. Extending this line of thinking we can draw on the 'engagement' literature and recognise that there is an opportunity to include the public in the design cycle. Whilst this is happening with a number of trials around the UK, raising awareness of these trials and grounding the 'when and where' in actual lived activity will empower the public to contribute.

**Knowledge from/ Trust with**

- **Experience**: Given the current development stage of self-driving technology less is known about the knowledge acquisition journey through usage. A small proportion of the public have experienced self-driving technology through a trial or in a simulator and data from these studies underpin much of this paper. We can also conceive 'trial' in a wider context in which an individual conducts their own trial-use of a new self-driving technology product or service. This is much the same as an individual deciding to have a ride on the new Jubilee Underground line in their spare time, before determining if you would use it every day on your commute. Such trial potentials are some way off but are nevertheless an important step on the engagement and trust building journey [92].
- **Using:** This represents the transition from knowledge acquisition to actual use in the context of an individual's daily life, which will come as self-driving technology mature. We can conceive this as not just using self-driving technology but also as interacting with them i.e. cyclists and pedestrians [94]. Such a phase in the engagement processes also relates to wider transport needs and habit formation. We can see a transition here from the (perceived) benefits stage above to realised benefits through usage, resulting in a feedback loop. Efforts should be made to make this a validation feedback loop, not a disproving one.
- **Malfunction:** Once in scaled operation malfunctions are inevitable**.** Public's awareness or attitude towards the limitations of self-driving technology and their subsequent ability to adapt their use of the system to accommodate whilst still deriving the expected benefits will play a significant role in how such malfunctions impact trust [4]. Consequently, there is a 'necessity of regularly calibrating the public's knowledge and expectation of autonomous vehicles through educational campaigns and legislative measures' [4]. Such in-service trust factors will be heavily influenced by the individual product or service design and associated cultures such as Brand trust.
- **Enforcement bodies:** The system-of-systems in which safety-critical technology establishes, is by definition, multifaceted. Trustworthy behaviour therefore cannot be established by a given individual or organisation, but as a defining characteristic of the system. Tangible characteristics of such a system include the availability of data sets, meeting munities and engaging with stakeholders [6]. The processes of making such information available underpins trust as well as engagement with that data itself. Seeing such ethical and regulatory consideration enforced is an area of interest to the public [38].

Recommendations: A distinct limitation identified in the literature is that the vast majority of studies, and by extension the public themselves, are yet to experience self-driving technology (even those that have do so only in a 'trial' context) [92]. Understanding that trust creation begins much before the first interactions between a user and a self-driving technology [13] we must refine our engagement approach to those that want, need, or would benefit from engagement at this time [12]. Whilst lived experience of self-driving technology will be an important component of engagement going forward, a primary focus on the trust *in* elements of the model is required at this stage of wide scale public engagement.

With a focus on the trust *in* elements of the model the creation of Proof-of-Concept engagement material that addresses each attribute within it is needed. This material will be used to facilitate engagements, learning and refinement and we define a process for doing so in the next section.

## The creation of engaging awareness and education material

Before we scale engagement material, that material must first be created. The process of creation has a substantial influence on the ability of that material to both engage the public, and scale. Our paper has found important learnings in stakeholder engagement and co-creation, both of which build on the principals previously identified, to guide this process. Before proposing a model for content creation that draws on this learning, we summarise these findings.

Stakeholder engagement is a critical aspect of effective engagement, continued and updated understanding and involving of stakeholders is essential [15]. Extending to a complex safety critical eco-system the role of stakeholder engagement includes addressing dynamic relations and complex environments that utilises different frames of reference, networks, specialist insights, and ongoing changes [15]. As such a stage of analysis should be undertaken to consider networks within which a stakeholder exists, their relative power and influence and their frames of reference [15]

Programmes of trust building are more successful when intended efforts to 'level the social and cognitive authority' of experts are implemented [7]. The public have a deficit of knowledge and that lack of knowledge can drive rejection [6]. Effective engagement must account for these imbalances and make efforts to address them.

A recurring theme in response to these points within the engagement literature is co-creation. Co-creation is characterised by the three Co's Framework of: Co-Define, Co-Design and Co-Refine. Co-creation facilitates people to harness the power they already have [62].

Drawing on these findings the model in Figure 7: Model for creating engaging content emerges. This model creates mechanisms to ensure engagement material remains factual and grounded, but for the actual material itself to undergo a process of co-creation. A key factor of importance to recognise is that the inclusion of key stakeholders at the start and throughout these processes will lead to those stakeholders themselves getting behind the final content. Stakeholders will have a stake in the result and actively 'booster' its distribution and further engagement.

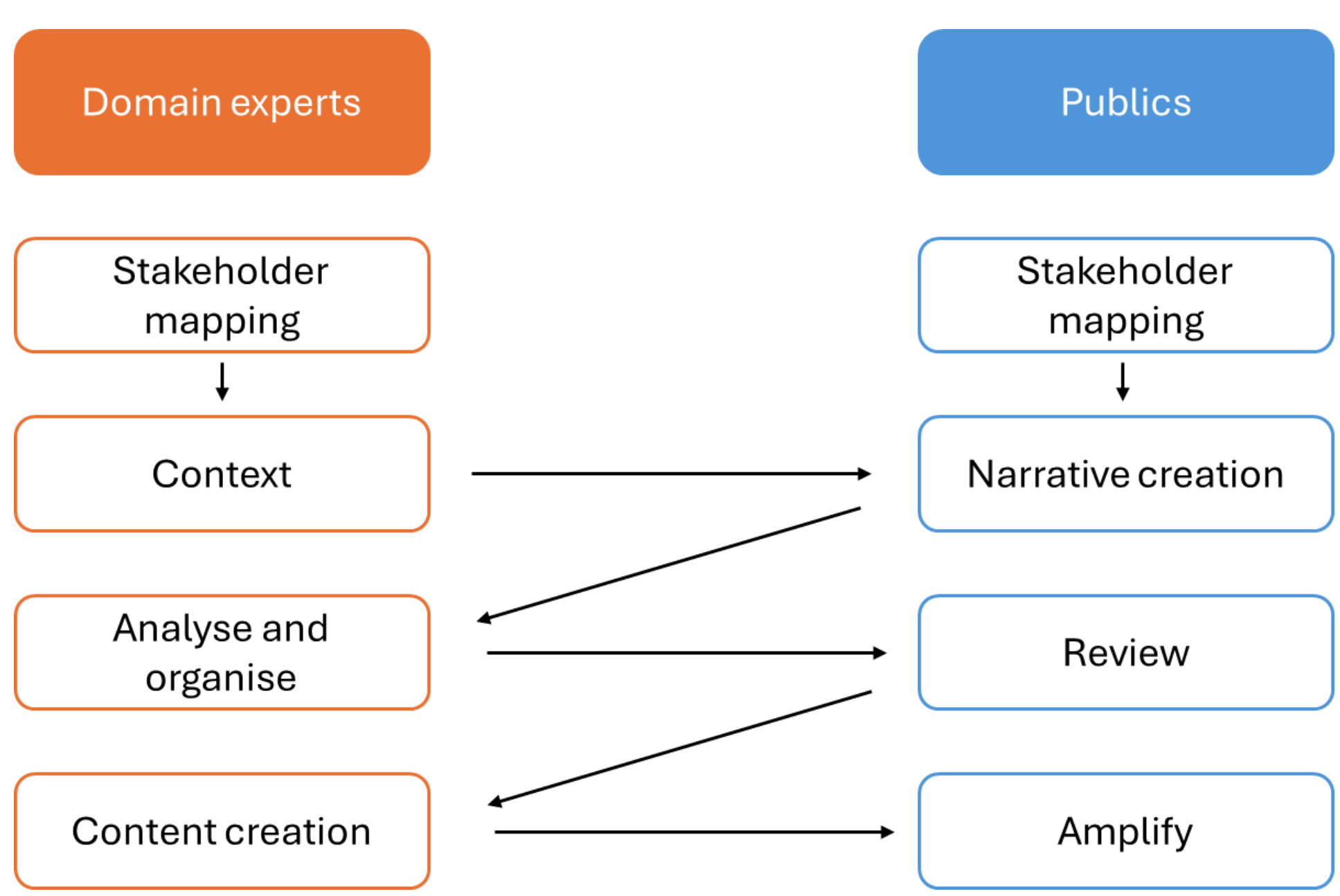

*Figure 7: Model for creating engaging content*

## Recommendations: Firstly, a thorough stakeholder mapping exercise is required, including an initial triage assessment of individual stakeholder relevance (contribution, need, want, would benefit from engagement), their relative knowledge deficit, their relative power to affect self-driving technology change, and their points of reference. These processes will benefit greatly from the personas work already undertaken in the UK [59] [60] and self-driving technology supply chain mapping [61]. Following stakeholder mapping stakeholders will be prioritised for co-creation engagement. A series of workshops should then be undertaken using the PoC material developed in the prior step to progress through the relevant stages of the proposed model. The range of stakeholders engaged and the number of workshops will be influenced by time and resource. However, is it recommended that such activities repeat along an appropriate timeframe (i.e. annually) to mirror the dynamic nature of knowledge acquisition and trust.

## Scalability

To scale content, we must align with ideas of communication, marketing and Public Relations (PR). However, this must be done in alignment with the overarching engagement and trust building principles already identified. Through this paper we have found that over the past decade there has been growing consensus that communications and PR is shifting to engagement ([21], [37], [38]). Such a framing understands Communications and PR as stakeholders' cognitive involvement and community capacity building through relationship building [21] in response to the desire of stakeholders to be involved in more dialogic forms of engagement [47]. But how do we realise these outcomes at scale?

In the first instance a comprehensive research strategy should be formed. Research makes communications and PR activities strategic by ensuring that communication is specifically targeted to the public who want, need, or care about the information [12]. We can see this paper itself as initial research towards a programme of self-driving technology engagement and the models proposed in this final section the process of organising that research into implemental frameworks. As a result, we are increasing our understanding, but not to use this information to persuade, but as the base for the building of mutual understanding between the public and organisations [50].

Empowering the public in a trustworthy process is characterised in the opportunity for stakeholders to actively contribute and affect outcomes (not just be listened to). The challenge at scale is retaining two-way engagement and not progressively moving towards traditional communications or one-way engagement. Furthermore, as the journey evolves the characteristics of engagement changes. If well-constructed the public will progressively become more active (engaged), acting as informed critiques and advocates when highly engaged [8]

We therefore build a model for scaling engagement that recognises these factors and uses stakeholders to amplify engagement reach. The inclusion of measurable outcomes here is key to a full research cycle. Evaluation is a critical aspect of an engagement campaign such that intended outcomes can be benchmarked and further research data can be collected to underpin a cycle of continuous improvement [12]

| Outcome | Content | Measures |
|---|---|---|
| **Direct engagement** | Expert-to-the public workshops, Imaginaries of the future, concept introductions (none-assumptive) discussion starters, events and outreach, study design and implementation | Passive: N/A<br>Active: One-way communication opportunities<br>Engagement: Evolving content outcomes |
| | | |
| **Amplified engagement through stakeholders** | Tool kits, narrative crib sheets, case studies, data, FAQ's + response guidance, panel/ event speakers, | Passive: Stakeholder reference material<br>Active: Stakeholders use material<br>Engagement: feedback and co-developed material |
| | | |
| **Amplified engagement through media** | Media briefing kit (fact sheet, talking points, FAQ's), blogs, articles, interviews | Passive: Media referencing material<br>Active: Media using material<br>Engagement: feedback and co-developed material |
| | | |
| **Wide scale engagement** | Webinars, social media, Questionnaires, Website, concept introductions, Case Studies | Passive: Clicks & views<br>Active: Likes & shares<br>Engagement: Feedback, comments & data |

**Recommendations**: Scaling engagement is heavily reliant on the prior recommendations having been implemented. It is also reliant on continued engagement at all levels which will provide a pipeline of new appropriate material. The core content that is generated from these stages i.e. What is a self-driving technology? Can then be adapted to the specific content needs of each outcome. When adapted, the prior steps will provide confidence that the messaging, engagement techniques and perspectives have relevance to the public. This in turn enhances the potential for the public to become progressively more active, maturing to informed critiques and advocates when highly engaged [8]. Achieving such an engagement outcome at scale will truly bring the public into the creation of a trustworthy self-driving technology system.

## Conclusive summary

What have we learnt and how can it be applied? We set out at the beginning of this paper to identify how Government and industry can prepare for wide scale awareness of self-driving technology among the public. To bring the public on every step of the journey a theoretical grounding in trust has been established, leading to 'engagement' programmes as an effective mechanism through which to establish the public role in the building of a trustworthy system. Robustness is built into this approach through understanding what matters to the public, and acknowledging this is often different to what domain experts consider important. A bridge between the public and domain experts is built through initial engagements focussed around what matters to the public that then creates a platform for new depths to be accessed. At each stage domain experts should acknowledge and calibrate for their relative knowledge and power advantage to sustain effective engagement throughout this journey.

A conclusive summary of each phase of the paper is provided at the end of each section. Drawing on these conclusions we have arrived at a model of engagement trust for self-driving technology. Drawing on threads of our learning this model begins with an overall framework for engagement and trust, becoming progressively more concrete in its application by systematically organising the attributes that form the public engagement journey with self-driving technology together with a framework for co-creation, finally arriving at a model of scalable self-driving technology engagement across the stakeholder landscape. Each phase of this model provides recommendations for practitioners implementing self-driving technology engagement programmes. Throughout the paper its recognised that self-driving technology products and services are yet to be operational. This has implications for the stage of the engagement journey we are on and the dynamic nature of trust that will shift in response to evolving knowledge acquisition of the public. Despite this comprehensive paper on self-driving technology trust and engagement, the journey is just beginning.